\documentclass[useAMS,usenatbib,usegraphicx]{mn2e}
\title[compressible MHD]{Compressible Magnetohydrodynamic 
Turbulence: mode coupling, scaling relations, 
anisotropy, viscosity-damped regime, and astrophysical implications}
\author[Cho \& Lazarian]{Jungyeon Cho,$^1$\thanks{E-mail:
cho@astro.wisc.edu} and A. Lazarian$^2$\thanks{E-mail:
lazarian@astro.wisc.edu}    \\
Astronomy Department, University of Wisconsin-Madison, 475 N.Charter
St., Madison, WI 53706, USA}
\begin{document}

\date{MNRAS, accepted}

\maketitle

\begin{abstract}
We present numerical simulations and explore scalings and
anisotropy of compressible magnetohydrodynamic (MHD) turbulence.
Our study covers both gas pressure dominated (high $\beta$) and 
magnetic pressure dominated (low $\beta$) plasmas
at different Mach numbers. In addition, we present
results for superAlfvenic turbulence and discuss in what way it 
is similar to the subAlfvenic turbulence. We describe a 
technique of separating different magnetohydrodynamic (MHD) modes 
(slow, fast and Alfven) and apply it to our simulations.
We show that, for both high and low $\beta$ cases,
Alfven and slow modes reveal the Kolmogorov $k^{-5/3}$ spectrum  
and scale-dependent Goldreich-Sridhar anisotropy, while
fast modes exhibit $k^{-3/2}$ spectrum and isotropy.
We discuss the statistics of density fluctuations arising from 
MHD turbulence at different regimes. Our findings 
entail numerous astrophysical implications ranging from cosmic ray
propagation to gamma ray bursts and star formation.
In particular, we show that the rapid decay of turbulence reported
by earlier researchers is not related to compressibility and 
mode coupling in MHD turbulence. In addition, we show that magnetic 
field enhancements and density enhancements are marginally correlated.
Addressing the density structure of partially ionized interstellar gas
on AU scales, we show that the viscosity-damped regime of MHD turbulence that
we earlier reported for incompressible flows persists for
compressible turbulence and therefore may provide an explanation
for those mysterious structures.
\end{abstract}


\section{Introduction}

Astrophysical turbulence is ubiquitous and it 
holds the key to many astrophysical
processes (star formation, heating of the interstellar medium, 
properties of accretion disks, 
cosmic ray transport etc). Therefore understanding of turbulence is
a necessary requirement for making further
progress along any of those directions.
Unlike laboratory turbulence astrophysical turbulence is magnetized
and highly compressible.

Turbulence has been studied in the context of the interstellar
medium (ISM) and the solar wind. The ISM in the Milky Way and neighboring
 galaxies is known to be turbulent on scales ranging from AUs to
kpc (Armstrong, Rickett \& Spangler 1995; 
Stanimirovic \& Lazarian 2001; Deshpande et
al. 2000). The solar wind also exhibits  small-scale turbulence 
(Leamon et al.~1998). The measured statistics of 
fluctuations in the ISM and the solar wind is consistent with 
the Kolmogorov turbulence obtained for 
incompressible unmagnetized fluid. This surprising observational
evidence\footnote{The ambiguities of the data interpretation were
     frequently quoted to justify ignoring this fact. For 
     instance, electron density fluctuations discussed in Armstrong et al.
     (1995) provide only indirect evidence for Kolmogorov-type spectrum. 
     However,
     the solar wind observations are made {\it in situ} and more difficult
     to disregard. Moreover, a newly developed statistical technique 
     (Lazarian \& Pogosyan 2000) allowed us
     to measure Kolmogorov type spectrum
     of velocity fluctuations (see also Lazarian \& Esquivel 2003).} 
resulted in numerous attempts to use
Kolmogorov statistics for practical computations of astrophysical
quantities, e.g. cosmic ray scattering. We shall show below that in
most cases such sort of calculations brings {\it erroneous} results.

Why would we expect astrophysical fluids to be turbulent and how
can we study astrophysical turbulence?
A fluid of viscosity $\nu$ gets turbulent when the rate of viscous 
dissipation, which is  $\sim \nu/L^2$ at the energy injection scale $L$, 
is much smaller than
the energy transfer rate $\sim V_L/L$, where $V_L$ is the velocity dispersion
at the scale $L$. The ratio of the two rates is the Reynolds number 
$Re=V_LL/\nu$. In general, when $Re$ is larger than $10-100$
the system becomes turbulent. Chaotic structures develop gradually as 
$Re$ increases,
and those with $Re\sim10^3$ are appreciably less chaotic than those
with $Re\sim10^8$. Observed features such as star forming clouds are
very chaotic for $Re>10^8$. 
This not only ensures that the fluids are turbulent but
also makes it difficult to simulate the turbulence. The currently available
3D simulations for 512 cubes can have $Re$ up to $\sim O(10^3)$
and are limited by their grid sizes. 
Therefore, it is essential to find ``{\it scaling laws}" in order to
extrapolate numerical calculations ($Re \sim O(10^3)$) to
real astrophysical fluids ($Re>10^8$). 
We show below that even with its limited resolution, numerics is a great 
tool for {\it testing} scaling laws. 

Kolmogorov theory provides a scaling law for incompressible 
{\it non}-magnetized hydrodynamic turbulence (Kolmogorov 1941).
This law is true in the statistical sense and it provides a relation
between the relative velocity $v_l$ of fluid elements and their separation
$l$, namely, $v_l\sim l^{1/3}$.  An equivalent description is to 
express spectrum $E(k)$
as a function of wave number $k$ ($\sim 1/l$).
The two descriptions are related by $kE(k) \sim v_l^2$. The famous
Kolmogorov spectrum is  $E(k)\sim k^{-5/3}$. The applications of 
Kolmogorov theory range from engineering research to
meteorology (see Monin \& Yaglom 1975) but its astrophysical
applications are poorly justified.

Let us consider {\it incompressible} MHD turbulence first.
There have long been understanding that the MHD turbulence
is anisotropic\footnote{
     It is not possible to cite all the important papers in the area
     of MHD turbulence. An incomplete list of the references in a 
     recent review on the statistics of
     MHD turbulence by Cho, Lazarian \& Vishniac (2003a; 
     henceforth CLV03a) includes about two hundred entries.} 
(e.g. Shebalin et al.~1983). A substantial progress has been achieved
recently by Goldreich \& Sridhar (1995; hereafter GS95) who made an
ingenious prediction regarding relative motions parallel and
perpendicular to magnetic field {\bf B} for incompressible
MHD turbulence. The GS95 model envisages a Kolmogorov spectrum of velocity 
and a scale-dependent anisotropy (see below).
These relations have been confirmed 
numerically (Cho \& Vishniac 2000b; Maron \& Goldreich 2001;
Cho, Lazarian \& Vishniac 2002a, hereafter CLV02a; see also CLV03a); 
they are in good agreement with observed and inferred astrophysical spectra 
(see CLV03a). A remarkable fact revealed in CLV02a is that
fluid motions perpendicular to {\bf B} are {\it identical} to hydrodynamic
motions. This provides an essential physical insight into why
in some respects MHD turbulence and hydrodynamic turbulence are
similar, while in other respects they are different.

The GS95 model considered incompressible MHD, but the real ISM is 
{\it highly compressible}. 
Literature on the properties of compressible MHD is very rich (see CLV03a).
Back in 80's Higdon (1984) theoretically studied density fluctuations
in the interstellar MHD turbulence.
Matthaeus \& Brown (1988) studied nearly incompressible MHD at low Mach
number and Zank \& Matthaeus (1993) extended it. In an important paper
Matthaeus et al.~(1996) numerically
explored anisotropy of compressible MHD turbulence. However, those
papers do not provide universal scalings of the GS95 type.

Is it feasible to obtain scaling relations for the compressible MHD
turbulence?
Some hints about effects of compressibility can be inferred from 
the Goldreich \& Sridhar (GS95) seminal paper. 
A more focused discussion was
presented in the Lithwick \& Goldreich (2001) paper which deals with electron
density fluctuations in the gas pressure dominated plasma, 
i.e.  in high $\beta$ regime ($\beta\equiv P_{gas}/P_{mag}\gg 1$). 
Incompressible regime formally
corresponds to $\beta\rightarrow \infty$ and therefore it is natural
to expect that for $\beta\gg 1$ the GS95 picture would
persist. Lithwick \&
Goldreich (2001) also speculated that for low $\beta$ plasmas the GS95
scaling of slow modes may be applicable.
An important 
study of MHD modes in compressible low $\beta$ plasmas is given in
Cho \& Lazarian (2002; hereafter CL02) where we developed and tested
our technique of separating different MHD modes.

In this work, we provide a detailed study of mode coupling and scalings
of compressible (fast and slow) and Alfvenic modes in 
high $\beta$, intermediate, and low $\beta$ plasmas.
Our approach is complementary to that employed in direct numerical
simulations of astrophysical turbulence. In such simulations, e.g.
in those dealing with the 
interstellar medium (see Vazquez-Semadeni et al.~2000),
simulations of particular astrophysical objects, e.g. molecular clouds,
are attempted. These simulations provide synthetic maps that can be compared
with observations. Our goal is to obtain scaling laws that can also
be compared with observations. 
In \S2, we describe our approach to the problem including both
simple theoretical considerations/expectations
that motivate our study and the numerical technique that we employ.
In \S3, we present velocity spectra and anisotropies  
for high and low $\beta$ plasmas.
In \S4, we discuss scalings of density and magnetic field.
In \S5, we present the study of viscosity-damped regime of MHD turbulence
in compressible fluid. This study extends our earlier work 
(Cho, Lazarian, \& Vishniac 2002b, henceforth CLV02b) where
this regime was reported for incompressible flows.
In \S6, we discuss astrophysical implications of our results,
including the rate of MHD turbulence decay, relation between the
decay rate and compressibility, correlation of density and magnetic
field, and formation of density structures at AU scale. 
The summary is given in \S7.

\section{Our Approach}

\subsection{Theoretical Considerations}  \label{section_theoretical}

\begin{figure*}
  \includegraphics{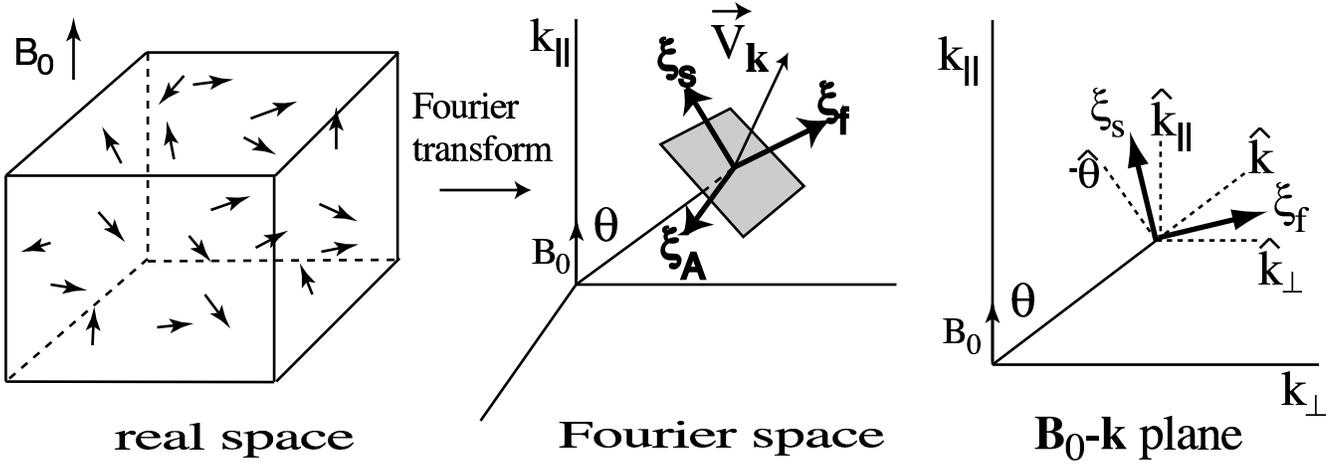}
  \caption{
      Separation method. We separate Alfven, slow, and fast modes in Fourier
      space by projecting the velocity Fourier component ${\bf v_k}$ onto
      bases ${\bf \xi}_A$, ${\bf \xi}_s$, and ${\bf \xi}_f$, respectively.
      Note that ${\bf \xi}_A = -\hat{\bf \varphi}$. 
      Slow basis ${\bf \xi}_s$ and fast basis ${\bf \xi}_f$ lie in the
      plane defined by ${\bf B}_0$ and ${\bf k}$.
      Slow basis ${\bf \xi}_s$ lies between $-\hat{\bf \theta}$ and 
      $\hat{\bf k}_{\|}$.
      Fast basis ${\bf \xi}_f$ lies between $\hat{\bf k}$ and 
      $\hat{\bf k}_{\perp}$.
}
\label{fig_separation}
\end{figure*}

Let us start with a discussion why isotropic Kolmogorov turbulence
{\it cannot} be applicable for describing strongly magnetized gas.
Assume that, at some large scale $L$, the magnetic energy and
kinetic energy are equal: $\rho V_L^2/2 \sim B^2/(4\pi)$.
According to the Kolmogorov theory, the kinetic energy at scale $l < L$ is
$\rho V_l^2/2 \sim (l/L)^{2/3} (\rho V_L^2/2)$, which is
smaller than the large scale kinetic energy by  a factor of $(l/L)^{2/3}$.
But, magnetic energy density does not diminish as the scale reduces.
Therefore, at scales smaller than $L$, hydrodynamic
motions will not be able to bend magnetic field lines substantially.

An important observation that leads to understanding of the GS95
scaling is that magnetic field cannot prevent mixing motions
of magnetic field lines if the motions
are perpendicular to the magnetic field. Those motions will cause, however,
waves that will propagate along magnetic field lines.
If that is the case, 
the time scale of the wave-like motions, i.e. $\sim l_{\|}/V_A$, where
$l_{\|}$ is the characteristic size of the perturbation along 
the magnetic field and 
$V_A=B/\sqrt{4 \pi \rho}$ is 
the local Alfven speed, will be equal to the hydrodynamic time-scale, 
$l_{\perp}/v_l$, 
where $l_{\perp}$ is the characteristic size of the perturbation
perpendicular to the magnetic field.
The mixing motions are 
hydrodynamic-like\footnote{
        Recent simulations (Cho et al.~2003) suggest that
        perpendicular mixing is indeed efficient for
        mean magnetic fields of up to the equipartition value.
        This corresponds to our earlier result that
        high order velocity statistics of MHD turbulence in the perpendicular
        directions is very similar to that of hydrodynamic one
        (CLV02a).
}
and therefore obey Kolmogorov scaling
$v_l\propto l_{\perp}^{1/3}$. Combining the two relations above
we can get the GS95 anisotropy, $l_{\|}\propto l_{\perp}^{2/3}$ 
(or $k_{\|}\propto k_{\perp}^{2/3}$ in terms of wave-numbers).
If  we interpret $l_{\|}$ as the eddy size in the direction of the 
local\footnote{The concept of {\it local} is crucial. The GS95
     scalings are obtained only in the local frame of magnetic field,
     as this is the frame where magnetic field are allowed to be
     mixed without being opposed by magnetic tension.} 
magnetic field
and $l_{\perp}$ as that in the perpendicular directions,
the relation implies that smaller eddies are more elongated
(see Appendix B for illustration of scale-dependent anisotropy).

How is this idealized incompressible model related to the actual,
e.g. interstellar, turbulence? 
Compressible MHD turbulence is a highly non-linear phenomenon
and it has been thought that 
different types of perturbations or modes (Alfven, slow and fast)
in compressible media are strongly coupled. Nevertheless,
one may question whether this is true.
A remarkable feature of the GS95 model is that
Alfven perturbations cascade to small scales over just one wave
period, while the other non-linear interactions require more time.
Therefore one might expect
that the non-linear interactions with other types of waves
should affect Alfvenic cascade only marginally. 
Moreover, as the Alfven waves are incompressible, the properties
of the corresponding cascade may not depend on the sonic Mach number.

The generation of compressible motions 
(i.e. {\it radial} components in Fourier space) 
{}from Alfvenic turbulence
is a measure of mode coupling.
How much energy in compressible motions is drained from Alfvenic cascade?
According to the closure calculations (Bertoglio, 
Bataille, \& Marion 2001; see also Zank \& Matthaeus 1993),
the energy in compressible modes in {\it hydrodynamic} turbulence scales
as $\sim M_s^2$ if $M_s<1$.
We may conjecture that this relation can be extended to MHD turbulence
if, instead of $M_s^2$, we use
$\sim (\delta V)_{A}^2/(a^2+V_A^2)$. 
(Hereinafter, we define $V_A\equiv B_0/\sqrt{4\pi\rho}$, where
$B_0$ is the mean magnetic field strength.) 
However, as the Alfven modes 
are anisotropic, 
this formula may require an additional factor.
The compressible modes are generated inside the so-called
Goldreich-Sridhar cone, which takes up $\sim (\delta V)_A/ V_A$ of
the wave vector space. The ratio of compressible to Alfvenic energy 
inside this cone is the ratio given above. 
If the generated fast modes become
isotropic (see below), the diffusion or, ``isotropization'' of the
fast wave energy in the wave vector space increase their energy by
a factor of $\sim V_A/(\delta V)_A$. This  results in
\begin{equation}
  \frac{ (\delta V)_{rad}^2 }{ (\delta V)_A^2 }    \la
 \left[ \frac{ V_A^2 + a^2 }{ (\delta V)^2_A } 
        \frac{ (\delta V)_A }{ V_A }   \right]^{-1},
\label{eq_high2}
\end{equation}
where $(\delta V)_{rad}^2$ and $(\delta V)_{A}^2$ are energy
of compressible\footnote{It is possible to show that 
the compressible modes inside the Goldreich-Sridhar cone
are basically fast modes.}  and Alfven modes, respectively.
Eq.~(\ref{eq_high2}) suggests that the drain of energy from
Alfvenic cascade is marginal when the amplitudes of perturbations
are weak, i.e. $(\delta V)_A \ll  V_A$.

If Alfven cascade evolves on its own, it is natural to assume that 
slow modes exhibit the GS95 scaling.
Indeed, slow modes in gas 
pressure dominated environment (high $\beta$ plasmas) are
similar to the pseudo-Alfven modes in incompressible regime 
(see GS95; Lithwick \& Goldreich 2001). The latter modes do follow
the GS95 scaling. 
In magnetic pressure dominated environments (low $\beta$ plasmas), 
slow modes are density perturbations propagating with the
sound speed $a$ parallel to the mean magnetic field
(see equation (\ref{xis_lowbeta})). Those perturbations are essentially
static for $a\ll V_A$. 
Therefore Alfvenic turbulence is expected to mix density
perturbations as if they were passive scalar. This also induces
GS95 spectrum.

The fast waves in low $\beta$ regime propagate at $V_A$ irrespectively
of the magnetic field direction. 
In high $\beta$ regime, the properties of fast modes are similar, 
but the propagation speed is the sound speed $a$.
Thus the mixing motions induced by Alfven waves should marginally
affect the fast wave
cascade. It is expected to
be analogous to the acoustic wave cascade and hence be isotropic.

For most part of this paper, we shall assumed that 
$\delta V \sim \delta B/\sqrt{4 \pi \rho} \sim B_0/\sqrt{4 \pi \rho}$,
where $\delta B$ is the r.m.s. strength of the random magnetic field.
This is less restrictive than it might appear, since as long as there is
some scale $l'$ in the turbulent cascade 
where $v_{l'}\sim B/\sqrt{4 \pi \rho}$ we
can take $L=l'$, $V_L=v_{l'}$ and use this model of turbulence
for all smaller scales.
        Suppose that we have a turbulent system initially threaded by a 
        mean magnetic field only.
If initially the turbulent energy is larger 
than the magnetic energy of the mean field, we are in the regime of 
so-called superAlfvenic turbulence. 
In this regime the growth of the magnetic 
field is expected through so called ``turbulent dynamo'' (see 
Batchelor 1950; Brandenburg et al. 1996; Cho \& Vishniac 2000a).
    Initially, the growth of magnetic energy is most active 
    at the scale roughly an order of magnitude larger 
    than the dissipation scale and
    the magnetic spectrum peaks at the scale.
    As the magnetic energy grows, 
    the magnetic back reaction becomes important at the
    scale 
    and the peak of the magnetic power spectrum 
    moves to larger scales.
    Finally, when equipartition between the kinetic and magnetic 
    energy densities occurs at a scale somewhat (factor of 2 or 3) 
    smaller than the kinetic energy peak, turbulence reaches
    a statistically stationary state. This agrees well
    with the results of incompressible simulations 
    in Cho \& Vishniac (2000a).
    However, this is not universally accepted idea.
    For example, Padoan \& Nordlund (1999) reported that,
    in their compressible simulations,
    the power spectrum of $\rho^{1/2}v$ has a shallower slope
    then those of velocity and magnetic field, which implies
    that equipartition between kinetic magnetic energy
    cannot be reached at small scales.
At scales smaller than the equipartition scale,
the turbulence becomes subAlfvenic and our earlier considerations
should be applicable.

The arguments above suggest that compressible MHD turbulence
should demonstrate well defined scaling relations. 
Below we test those arguments and reveal scaling relations for
different $\beta$s and Mach numbers.

\begin{figure*}
  \includegraphics[width=0.34\textwidth]{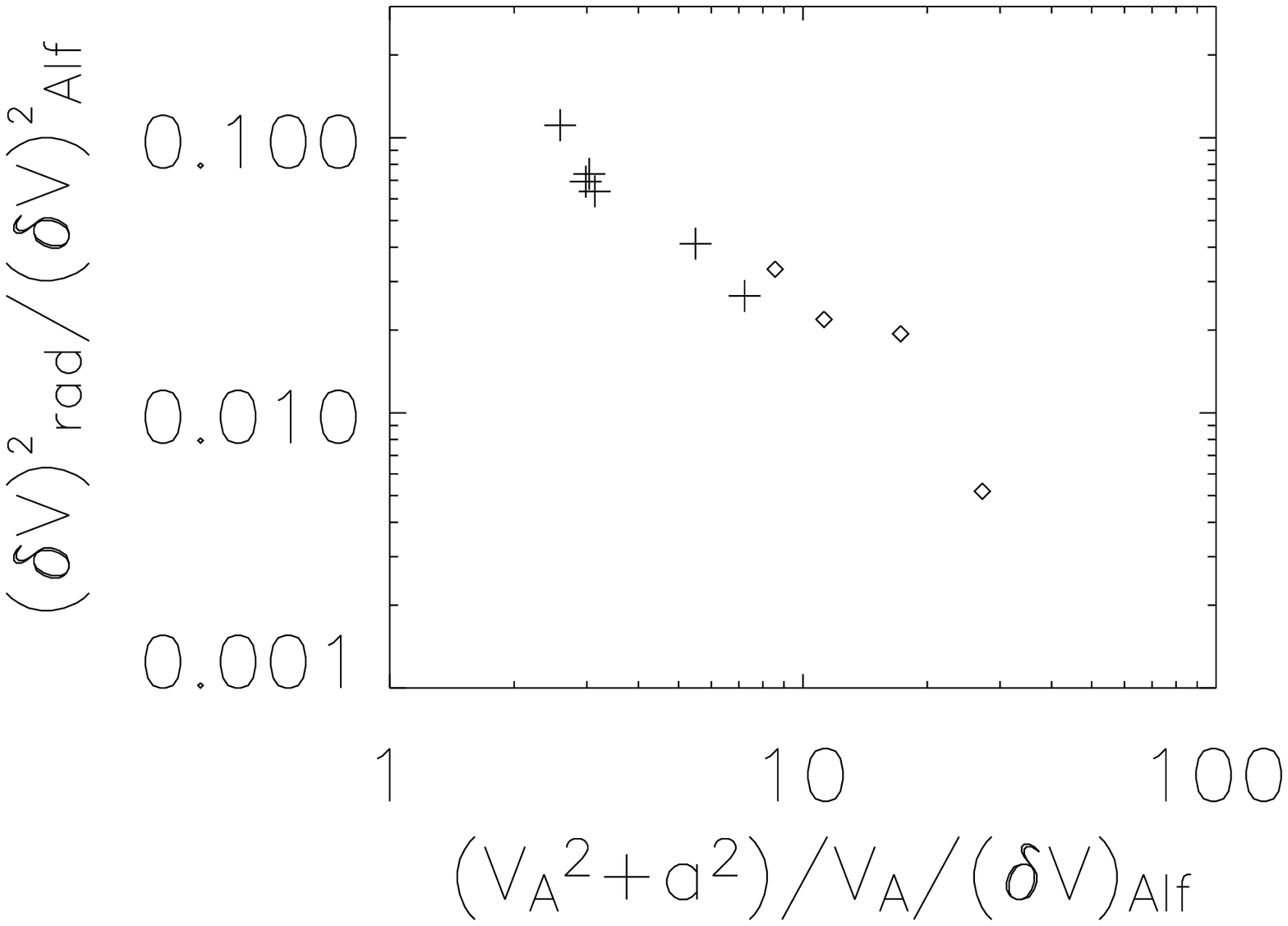}
\hfill
  \includegraphics[width=0.24\textwidth]{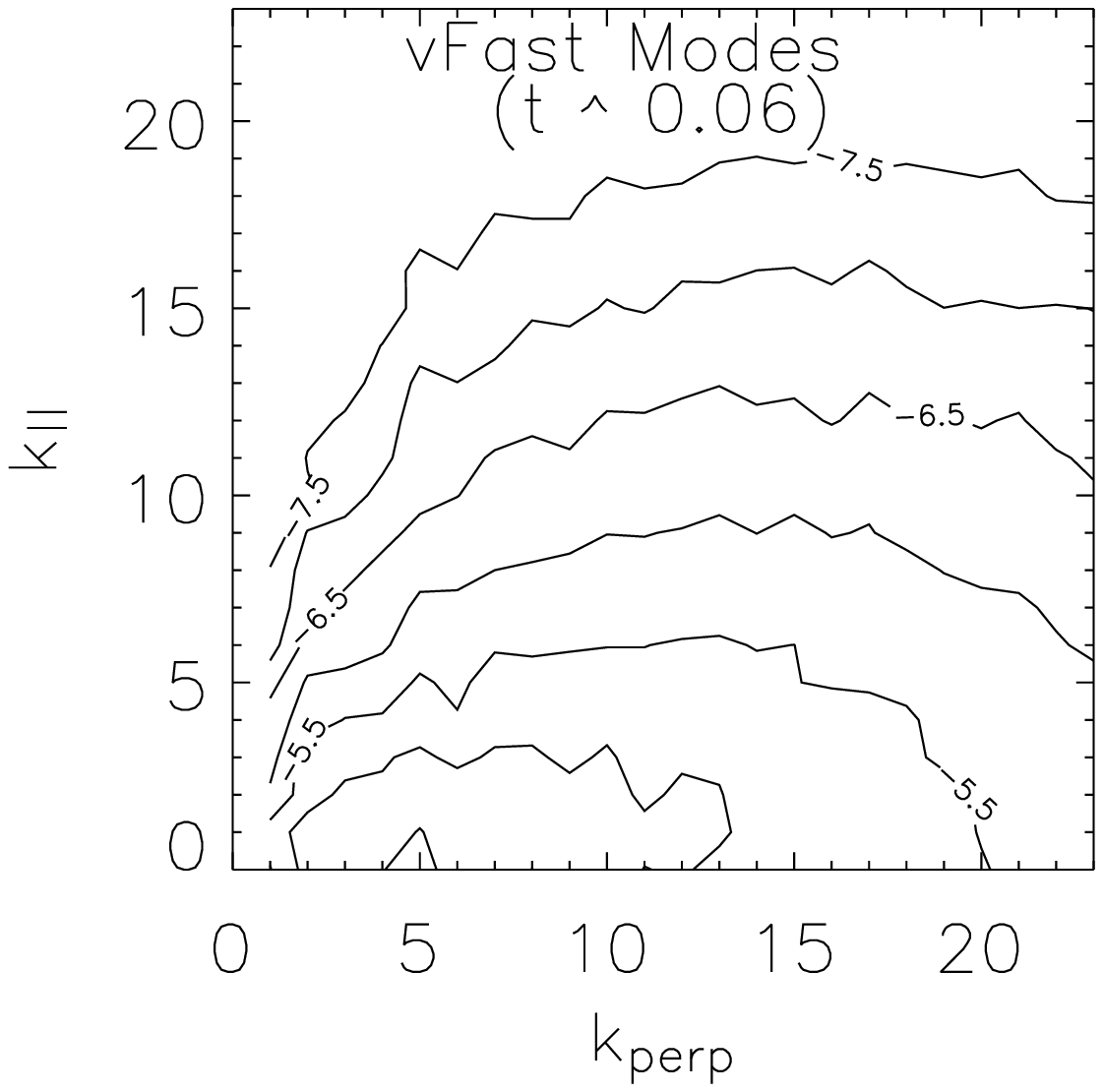}
\hfill
  \includegraphics[width=0.3\textwidth]{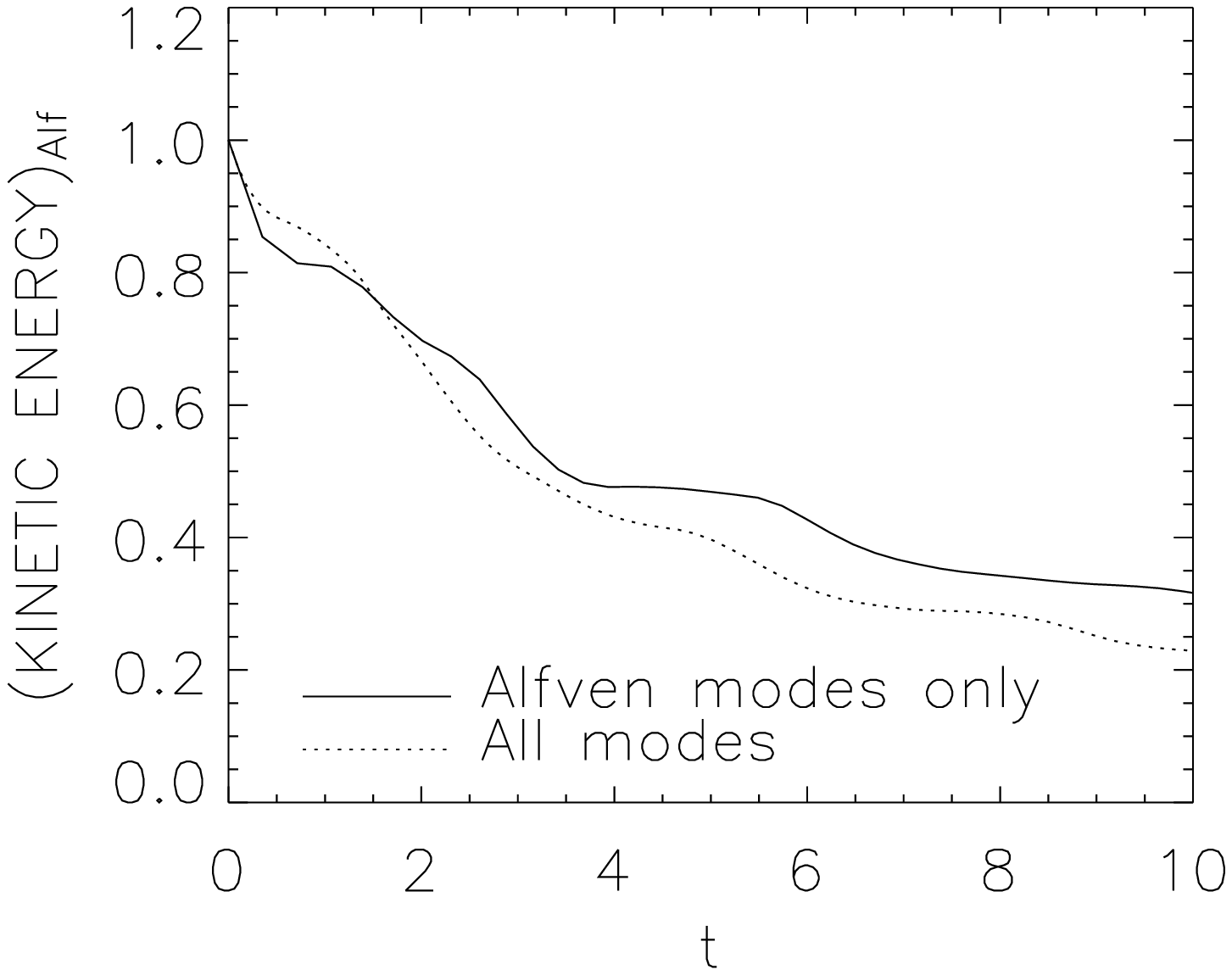}
  \caption{
      Mode coupling studies.
    (a){\it left:}  Square of the r.m.s. velocity of the compressible modes.
        We use $144^3$ grid points. Only Alfven modes are allowed
        as the initial condition.
        ``Pluses'' are for low $\beta$ cases ($0.02 \leq \beta \leq 0.4$).
        ``Diamonds'' are for high  $\beta$ cases ($1 \leq \beta \leq 20$).
    (b){\it middle:} Generation of fast modes. Snapshot is taken at t=0.06 from
        a simulation (with $144^3$ grid points) 
        that started off with Alfven modes only.
        Initially, $\beta$ (ratio of gas to magnetic pressure, $P_g/P_{mag}$) 
          $=0.2$ and 
          $M_s$ (sonic Mach number) $\sim 1.6$.
    (c){\it right:} Comparison of decay rates.
        Decay of Alfven modes is not much affected by other 
       (slow and fast) modes. We use $216^3$ grid points.
        Initially, $\beta=0.02$ and 
        $M_s\sim 4.5$ for the solid line and 
        $M_s\sim 7$ for the dotted line. 
        Note that initial data are, in some sense, identical for
        the solid and the dotted lines.
        The sonic Mach number for the solid line is smaller
        because we removed fast and slow modes from the initial data before
        the decay simulation.
        For the dotted line, we did {\it not} remove any modes from the
        initial data.
}
\label{fig_coupling}
\end{figure*}

\subsection{Numerical scheme}
We use a third-order accurate hybrid
essentially non-oscillatory (ENO) scheme (see CL02) 
to solve the ideal isothermal
MHD equations in a periodic box:
\begin{eqnarray}
{\partial \rho    }/{\partial t} + \nabla \cdot (\rho {\bf v}) =0,  \\
{\partial {\bf v} }/{\partial t} + {\bf v}\cdot \nabla {\bf v} 
   +  \rho^{-1}  \nabla(a^2\rho)
   - (\nabla \times {\bf B})\times {\bf B}/4\pi \rho ={\bf f},  \\
{\partial {\bf B}}/{\partial t} -
     \nabla \times ({\bf v} \times{\bf B}) =0, 
\end{eqnarray}
with
    $ \nabla \cdot {\bf B}= 0$ and an isothermal equation of state.
Here $\bf{f}$ is a random large-scale driving force, 
$\rho$ is density,
${\bf v}$ is the velocity,
and ${\bf B}$ is magnetic field.
The rms velocity $\delta V$ is maintained to be approximately unity
(in fact $\delta V \sim 0.7$), so that
 ${\bf v}$ can be viewed as the velocity 
measured in units of the r.m.s. velocity
of the system and ${\bf B}/\sqrt{4 \pi \rho}$ 
as the Alfv\'{e}n velocity in the same units.
The time $t$ is in units of the large eddy turnover time ($\sim L/\delta V$) 
and the length in units of $L$, the scale of the energy injection.
The magnetic field consists of the uniform background field and a
fluctuating field: ${\bf B}= {\bf B}_0 + {\bf b}$.

For mode coupling studies (Fig.~\ref{fig_coupling}), 
we do {\it not} drive turbulence.
For scaling studies, 
we drive turbulence solenoidally in Fourier space and
use $216^3$ points, $V_A=B_0/\sqrt{4 \pi \rho}=1$, and $\rho_0=1$. 
The average rms velocity in statistically stationary state is 
$\delta V\sim 0.7$.

For our calculations we assume that
$B_0/\sqrt{4 \pi \rho} \sim \delta B/\sqrt{4 \pi \rho} \sim \delta V$.
In this case, the sound speed is the controlling parameter and
basically two regimes can exist: supersonic and subsonic.
Note that supersonic means low-beta and subsonic means high-beta.
When supersonic, we consider mildly supersonic (or, mildly low-$\beta$)
and highly supersonic (or, very low-$\beta$)\footnote{
        The terms ``mildly'' and ``highly''  are 
        somewhat arbitrary terms.
        We consider these two supersonic cases to cover
        a broad range of parameter space.
        Note that Boldyrev, Nordlund, \& Padoan (2002b)
        recently provided a Mach number dependence study of the 
        compressible MHD turbulence statistics where only two regimes 
        are manifest: essentially incompressible and essentially 
        compressible shock-dominated (with smooth transition at some $M_s$ of 
        order unity).
}.

\subsection{Separation of MHD modes}    \label{section_decomposition}
Three types of waves exist (Alfven, slow and fast)
in compressible magnetized plasma. 
The slow, fast, and Alfven bases that denote the direction of displacement
vectors for each mode are given by
\begin{eqnarray}
   \hat{\bf \xi}_s \propto 
     ( -1 + \alpha - \sqrt{D} )
            k_{\|} \hat{\bf k}_{\|} 
     + 
     ( 1+\alpha - \sqrt{D} ) k_{\perp} \hat{\bf k}_{\perp},
  \label{eq_xis_new}
\\
   \hat{\bf \xi}_f \propto 
     ( -1 + \alpha + \sqrt{D} )
           k_{\|}  \hat{\bf k}_{\|} 
     + 
     ( 1+\alpha + \sqrt{D} ) k_{\perp} \hat{\bf k}_{\perp},  
   \label{eq_xif_new}
\\
 \hat{\bf \xi}_A = -\hat{\bf \varphi} 
         = \hat{\bf k}_{\perp} \times \hat{\bf k}_{\|},
\end{eqnarray}
where $D=(1+\alpha)^2-4\alpha \cos\theta$, $\alpha=a^2/V_A^2=\beta(\gamma/2)$,
$\theta$ is the angle between ${\bf k}$ and ${\bf B}_0$, and
$\hat{\bf \varphi}$ is the azimuthal basis in the spherical polar coordinate
system.
These are equivalent to the expression in CL02:
\begin{eqnarray}
   \hat{\bf \xi}_s &\propto &
        k_{\|} \hat{\bf k}_{\|}+
     \frac{ 1-\sqrt{D}-{\beta}/2  }{ 1+\sqrt{D}+{\beta}/2  } 
    \left[ \frac{ k_{\|} }{ k_{\perp} }  \right]^2
     k_{\perp} \hat{\bf k}_{\perp},  \label{eq_xis}     \\
   \hat{\bf \xi}_f &\propto &
     \frac{ 1-\sqrt{D}+{\beta}/2  }{ 1+\sqrt{D}-{\beta}/2  } 
    \left[ \frac{ k_{\perp} }{ k_{\|} } \right]^2
     k_{\|} \hat{\bf k}_{\|}  +
          k_{\perp} \hat{\bf k}_{\perp}.
\end{eqnarray}
(Note that $\gamma=1$ for isothermal case.)

Slow and fast velocity components can be obtained 
by projecting velocity Fourier component 
${\bf v}_{\bf k}$ onto $\hat{\bf \xi}_s$ and $\hat{\bf \xi}_f$, respectively.
In Appendix~A, we discuss how to separate slow and fast magnetic modes.
We obtain energy spectra using this projection method.
When we calculate 
the structure functions (e.g.~for Alfv\'{e}n velocity)
in real space, 
we first obtain the Fourier components using the projection and, then, 
we obtain the real space values by performing Fourier transform.


\begin{figure*}
  \includegraphics{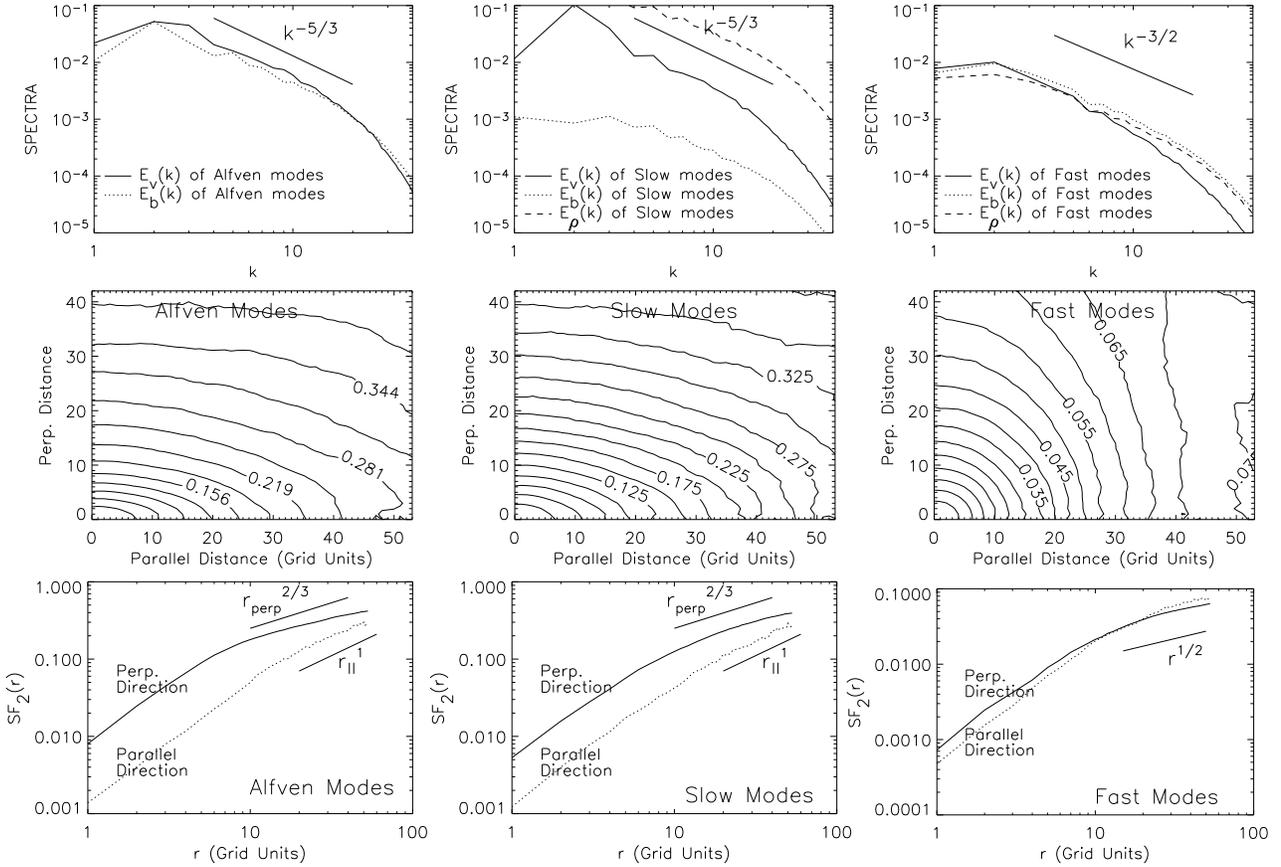}
  \caption{
      Low $\beta$ ($\beta\sim 0.2$ and $M_s\sim$2.3). 
     Scalings relations. Results from driven turbulence with
          $M_A$ (Alfven Mach number) $\sim 0.7$, 
          and $216^3$ grid points. ($V_A\equiv B_0/\sqrt{4\pi\rho}=1$; 
           $a$ (sound speed) $=\sqrt{0.1}$; 
          $\delta V \sim 0.7$.)
         (a){\it Upper-left:} 
             Spectra of Alfv\'en modes follow a Kolmogorov-like power law.
         (b){\it Middle-left:} The second-order structure function 
             ($SF_2$) 
             for velocity of Alfv\'en modes
             shows anisotropy similar to the GS95
        ($r_{\|}\propto r_{\perp}^{2/3}$ or $k_{\|}\propto k_{\perp}^{2/3}$).
         The structure functions are measured in directions perpendicular or
             parallel to the local mean magnetic field in real space.
             We obtain real-space velocity and magnetic fields 
             by inverse Fourier transform of
             the projected fields.
         (c){\it Lower-left:} SF$_2$ on the parallel axis 
            and on perpendicular axis for Alfven modes velocity.
         (d){\it Upper-middle:}  
            Spectra of slow modes also follow a Kolmogorov-like power law.
         (e){\it Middle-middle:} 
             Slow mode velocity shows anisotropy similar to the GS95.
         (f){\it Lower-middle:} SF$_2$ on the parallel axis 
            and on perpendicular axis for slow modes velocity.
         (g){\it Upper-right:} Spectra of fast modes are compatible with
             the IK spectrum.
         (h){\it Upper-middle:} The SF$_2$ of 
             fast modes velocity shows isotropy.
             Fast mode magnetic field also shows isotropy.
         (i){\it Lower-right:} SF$_2$ on the parallel axis 
            and on perpendicular axis for fast modes velocity.
}
\label{fig_M2}
\end{figure*}

\section{Velocity scalings}

\subsection{Mode coupling}

\begin{figure*}
  \includegraphics[width=0.99\textwidth]{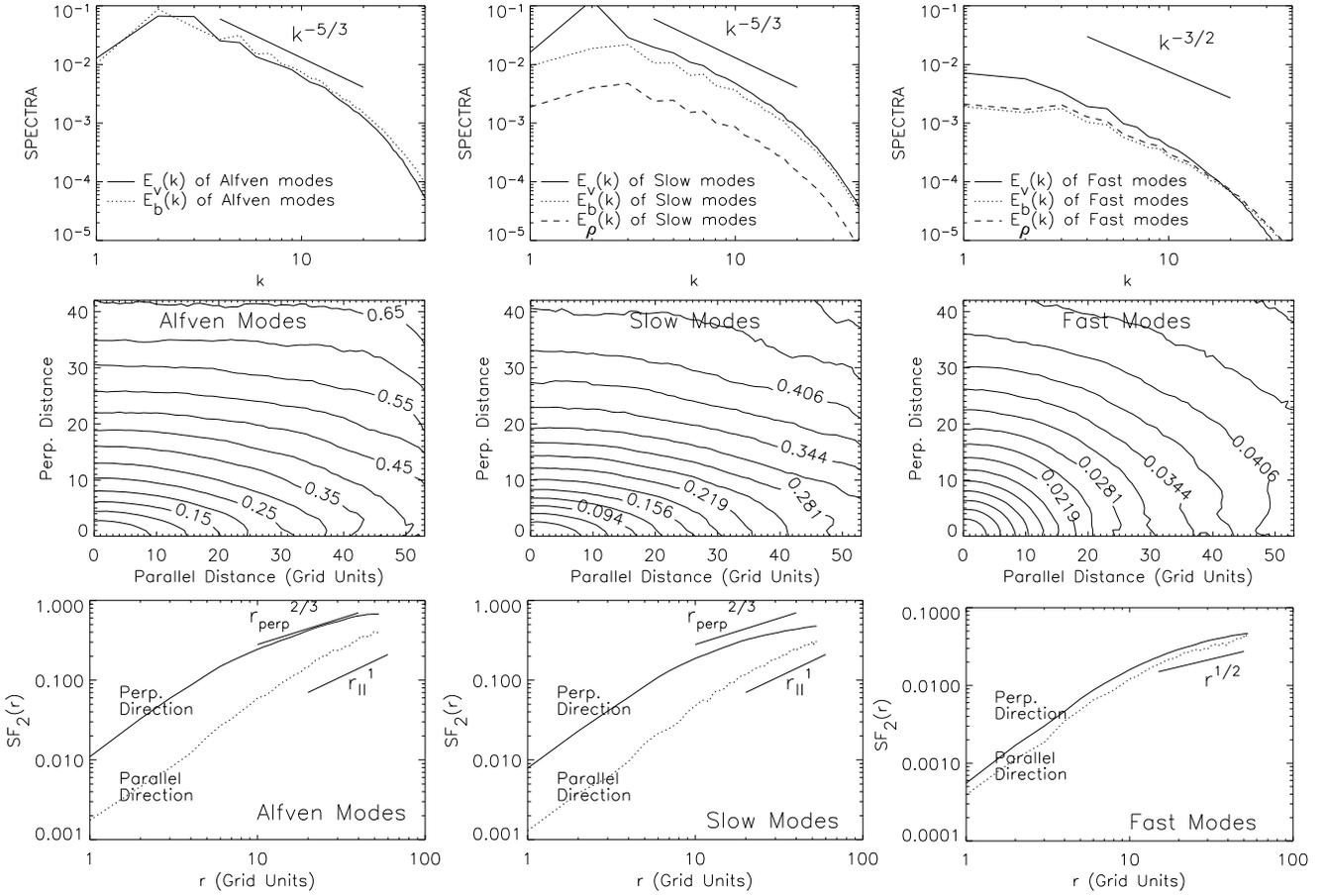}
  \caption{
      High $\beta$ ($\beta\sim 4$ and $M_s \sim$0.35). 
      $V_A\equiv B_0/\sqrt{4\pi\rho}=1$. $a$ (sound speed) $=\sqrt{2}$. 
      $\delta V \sim 0.7$.
      See caption in Fig.~\ref{fig_M2}.
      Alfven and slow modes follow the GS95 scalings.
      Fast modes are isotropic.
}
\label{fig_HB}
\end{figure*}

In CL02 we demonstrated the decoupling of Alfven and fast modes 
in low $\beta$ plasmas. Here we substantially extend the CL02 analysis.
As mentioned above, the coupling of compressible and incompressible modes
is crucial. If Alfvenic modes produce
a copious amount of compressible modes, the whole picture of independent
Alfvenic turbulence fails. However, our calculations  show
that the amount of energy drained into compressible motions is negligible,
provided that either the external magnetic field or
the gas pressure is sufficiently high.
Fig.~\ref{fig_coupling}a suggests that the generation of
compressible motions follows equation (\ref{eq_high2}).
Fast modes also follow a similar scaling, although the scatter is a bit larger.
        The marginal generation of compressible 
        modes is in agreement with 
        earlier studies by Boldyrev et al. (2002b) and 
        Porter, Pouquet, \& Woodward (2002),
        where the
        velocity was decomposed into a potential component
        and a solenoidal component.
        See Fig.~\ref{fig_coupling}a for the values of  $\sim \chi$
     (=the ratio of the mean square potential to solenoidal velocity).
Fig.~\ref{fig_coupling}b demonstrates that fast modes are initially generated
anisotropically, which supports our theoretical consideration 
in \S\ref{section_theoretical}.

Fig.~\ref{fig_coupling}c shows that dynamics of Alfven modes is 
not affected by slow modes.
       We first perform a driven turbulence simulation with $216^3$ 
       grid point, $\beta \sim 0.02$, and $M_s \sim 7$. 
       Then, after it has reached a statistically stationary
       state, we stop the run and save the data.
       Using these data, we perform two decay simulations.
       For one (the solid line), we remove all slow and fast modes and
       let the turbulence decay.
       For the other (the dotted line), {\it without} removing any modes,
       we just let the turbulence decay.
The solid line in the figure is the energy in Alfven modes
when we start the decay simulation with Alfven modes only.
The dotted line is the Alfven energy when we start the simulation with
all modes. 
This result confirms that Alfven modes cascade is 
almost independent of slow and fast modes.
In this sense, coupling between Alfven and other modes is weak.

\subsection{High-$\beta$ and
         mildly supersonic low-$\beta$ regimes}

\noindent 
{\bf Alfv\'{e}n Modes.---} 
{}Fig.~\ref{fig_M2}a (low-$\beta$) and Fig.~\ref{fig_HB}a (high-$\beta$)
show that the power spectra of Alfv\'{e}n waves follow
a Kolmogorov spectrum:
\begin{equation}
 \mbox{\it Spectrum of Alfv\'{e}n Waves:~~~}  
     E^{A}(k) \propto k_{\perp}^{-5/3}.
\end{equation}
In Fig.~\ref{fig_M2}b ({\it middle-left} panel) 
and Fig.~\ref{fig_HB}b ({\it middle-left} panel), 
we plot contours of equal second-order structure function for velocity
($SF_2({\bf r})=<|{\bf v}({\bf x}+{\bf r}) - 
                 {\bf v}({\bf x})|^2>_{avg.~over~{\bf x}}$)
obtained in local coordinate systems in which the parallel axis is aligned
with the local mean field (see Cho \& Vishniac 2000b; CLV02a;
Maron \& Goldreich 2001).
The $SF_2$
 along the axis perpendicular to the local mean magnetic field
follows a scaling compatible with $r^{2/3}$.
The $SF_2$ along the axis parallel  to the local mean field follows
steeper $r^{1}$ scaling 
(Fig.~\ref{fig_M2}c and Fig.~\ref{fig_HB}c: {\it lower-left} panels).
The results show reasonable agreement with the GS95 model  for incompressible 
MHD turbulence,
\begin{equation}
  \mbox{\it Anisotropy of Alfv\'{e}n Waves:~~~}  
  r_{\|}\propto r_{\perp}^{2/3}, \mbox{~or~} k_{\|}\propto k_{\perp}^{2/3},
\end{equation}
where
$r_{\|}$ and $r_{\perp}$ are the semi-major axis and semi-minor axis of eddies,
respectively (Cho \& Vishniac 2000b).

\vspace{0.3cm}
\noindent
{\bf Slow waves.---}
The incompressible limit of slow waves is pseudo-Alfv\'{e}n waves.
Goldreich \& Sridhar (1997) argued that
the pseudo-Alfv\'{e}n waves are slaved to the shear-Alfv\'{e}n 
(i.e.~ordinary Alfv\'{e}n)
waves, which
means that
pseudo-Alfv\'{e}n modes do not cascade energy for themselves. 
Lithwick \& Goldreich (2001) made similar theoretical arguments 
for high $\beta$ plasmas and
conjectured similar behaviors of slow modes in low $\beta$ plasmas.
We confirmed that similar arguments are also applicable to slow waves
in low $\beta$ plasmas (CL02).
Indeed, power spectra in 
Fig.~\ref{fig_M2}d and Fig.~\ref{fig_HB}d  
({\it upper-middle} panels) are consistent with:
\begin{equation}
 \mbox{\it Spectrum of Slow Modes:~~~}  E^{s}(k) \propto k_{\perp}^{-5/3}.
\end{equation}
In Fig.~\ref{fig_M2}e and Fig.~\ref{fig_HB}e ({\it middle-middle} panels),
contours of equal second-order 
velocity structure function ($SF_2$), 
representing eddy shapes,
show scale-dependent anisotropy: smaller eddies are more elongated.
The results show reasonable agreement with the GS95-type anisotropy 
\begin{equation}
\mbox{\it Anisotropy of Slow Modes:~~~}  
  k_{\|}\propto k_{\perp}^{2/3}, \mbox{~or~} 
  r_{\|}\propto r_{\perp}^{2/3}, 
\end{equation}
where
$r_{\|}$ and $r_{\perp}$ are the semi-major axis and semi-minor axis of eddies,
respectively.

\begin{figure*}
  \includegraphics[width=0.99\textwidth]{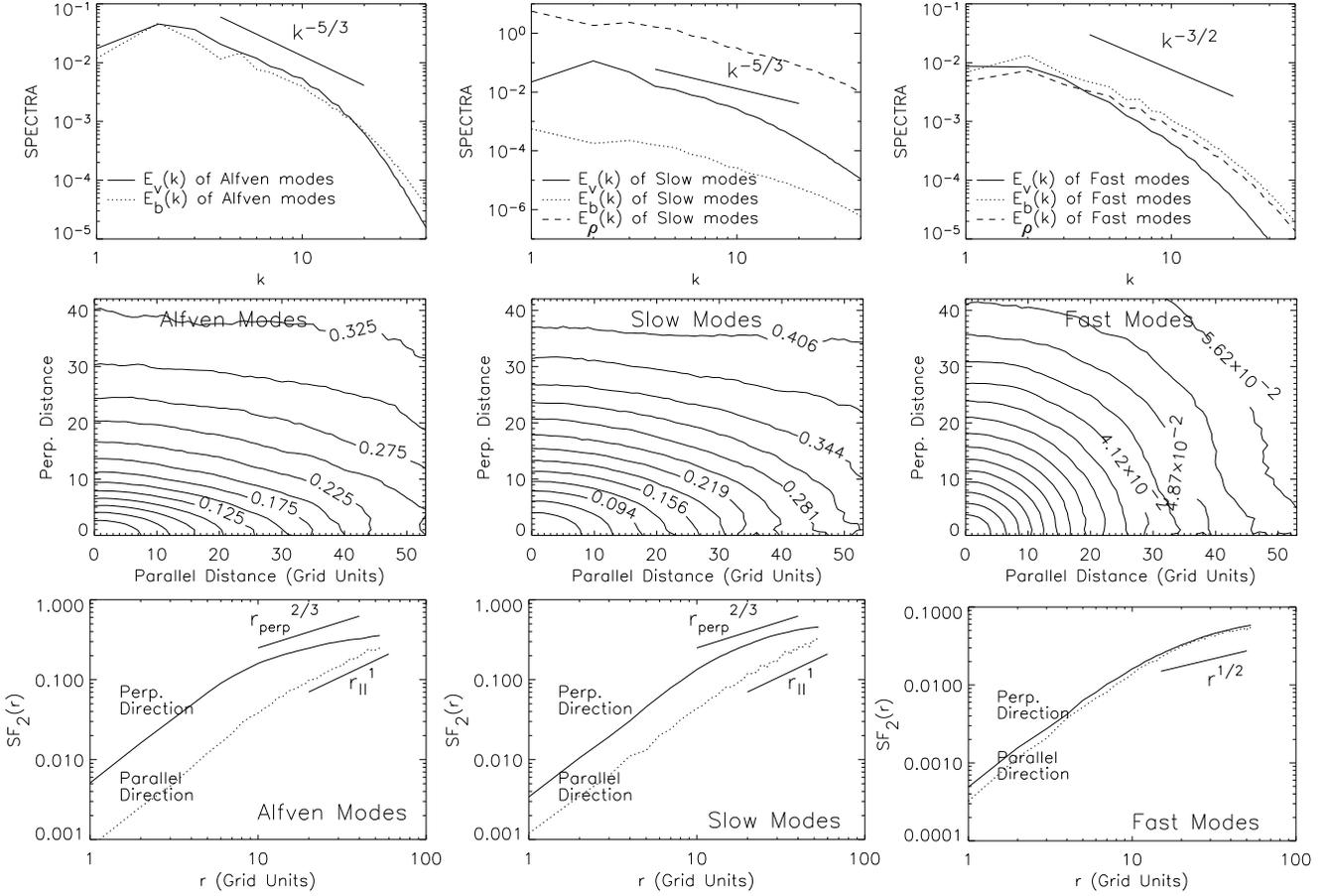}
  \caption{
      Highly supersonic low $\beta$ ($\beta\sim0.02$ and $M_s\sim$7). 
      $V_A\equiv B_0/\sqrt{4\pi\rho}=1$. $a$ (sound speed) $=0.1$. 
      $\delta V \sim 0.7$.
      Alfven modes follow the GS95 scalings. Slow modes follow
      the GS95 anisotropy. But velocity spectrum of slow modes is uncertain.
      Fast modes are isotropic.
}
\label{fig_M10}
\end{figure*}

\vspace{0.3cm}
\noindent
{\bf Fast waves.---}
{}Fig.~\ref{fig_M2}h and Fig.~\ref{fig_HB}h ({\it middle-right} panels)
show fast modes are isotropic.
{}The resonance conditions for the interacting fast waves are
$ 
\omega_1 + \omega_2 = \omega_3 \mbox{~~and~~}
  {\bf k}_1 + {\bf k}_2 = {\bf k}_3.
$ 
Since $ \omega \propto k$ for the fast modes,
the resonance conditions can be met only when
all three ${\bf k}$ vectors are collinear.
This means that the direction of energy cascade is 
{\it radial} in Fourier space.
This is very similar to acoustic turbulence, turbulence caused by interacting
sound waves (Zakharov 1967; Zakharov \& Sagdeev 1970; 
                  L'vov, L'vov, \& Pomyalov 2000).
Zakharov \& Sagdeev (1970) found
$E(k)\propto k^{-3/2}$.
However, there is debate about
the exact scaling of acoustic turbulence.
Here we cautiously claim that our numerical results are compatible
with the Zakharov \& Sagdeev scaling:
\begin{equation}
\mbox{\it Spectrum of Fast Modes:~~~} E^f(k) \sim  k^{-3/2}.
\end{equation}
The eddies are isotropic (see also Appendix B).

\subsection{Highly supersonic low-$\beta$ case}
The results for low-$\beta$ 
in the previous subsection are for a Mach number of $\sim 2.3$.
In this subsection, we present results for a Mach number of $\sim 7$.
Obviously shock formation is faster when the Mach number of the system is high.
We also expect that turbulent motions can compress/disperse the gas
more easily when the Mach number is high.
As a result, we expect higher density fluctuations when Mach number is higher.
Thus we check the scaling relations
for high Mach number fluids.

Fig. \ref{fig_M10} shows that most of the scaling relations that hold
true in mildly supersonic flows are still valid
in the highly supersonic case.
Especially anisotropy of Alfven, slow, and fast modes
is almost identical to the one in the previous section.
However, the power spectra for slow modes do not show
the Kolmogorov slope.
The slope is close to $-2$, which is suggestive of shock formation.
At this moment, it is not clear whether or not the $-2$ slope
is the true slope.
In other words, the observed $-2$ slope might be due to the limited
numerical resolution.
Runs with higher numerical resolution should
give the definite answer.

\subsection{SuperAlfvenic turbulence}  \label{sect_super}
\begin{figure*}
  \includegraphics[width=0.35\textwidth]{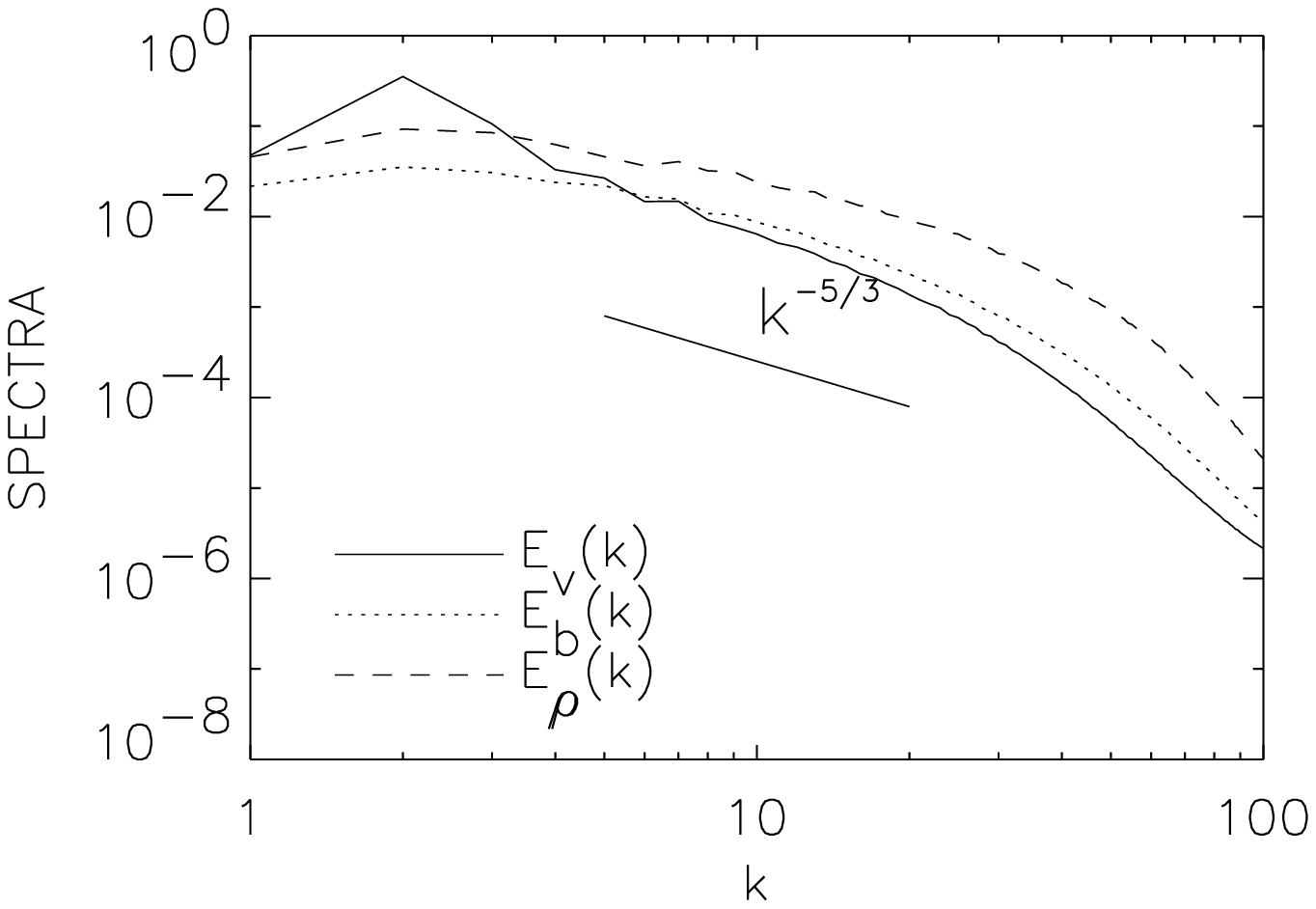}
\hfill
  \includegraphics[width=0.32\textwidth]{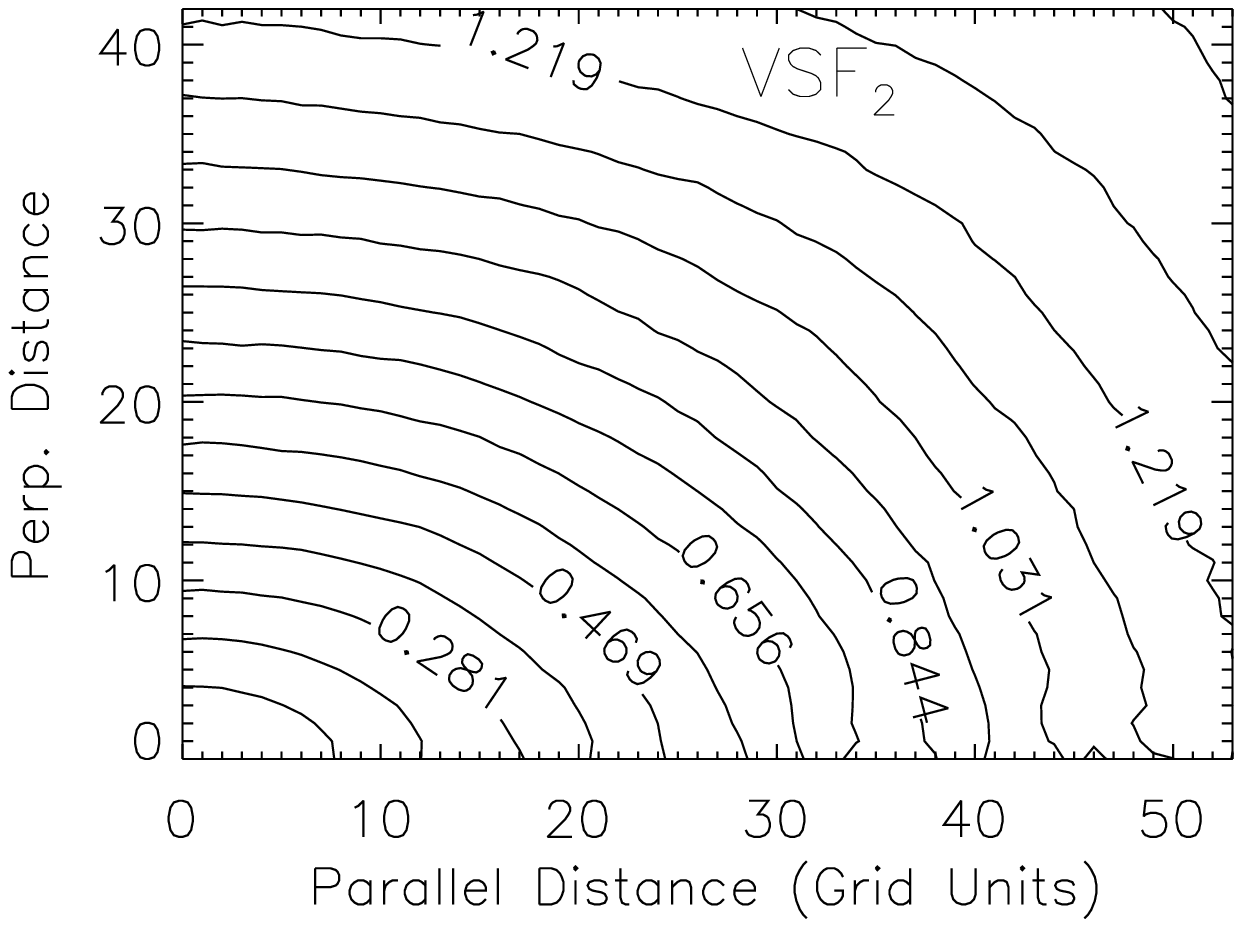}
\hfill
  \includegraphics[width=0.32\textwidth]{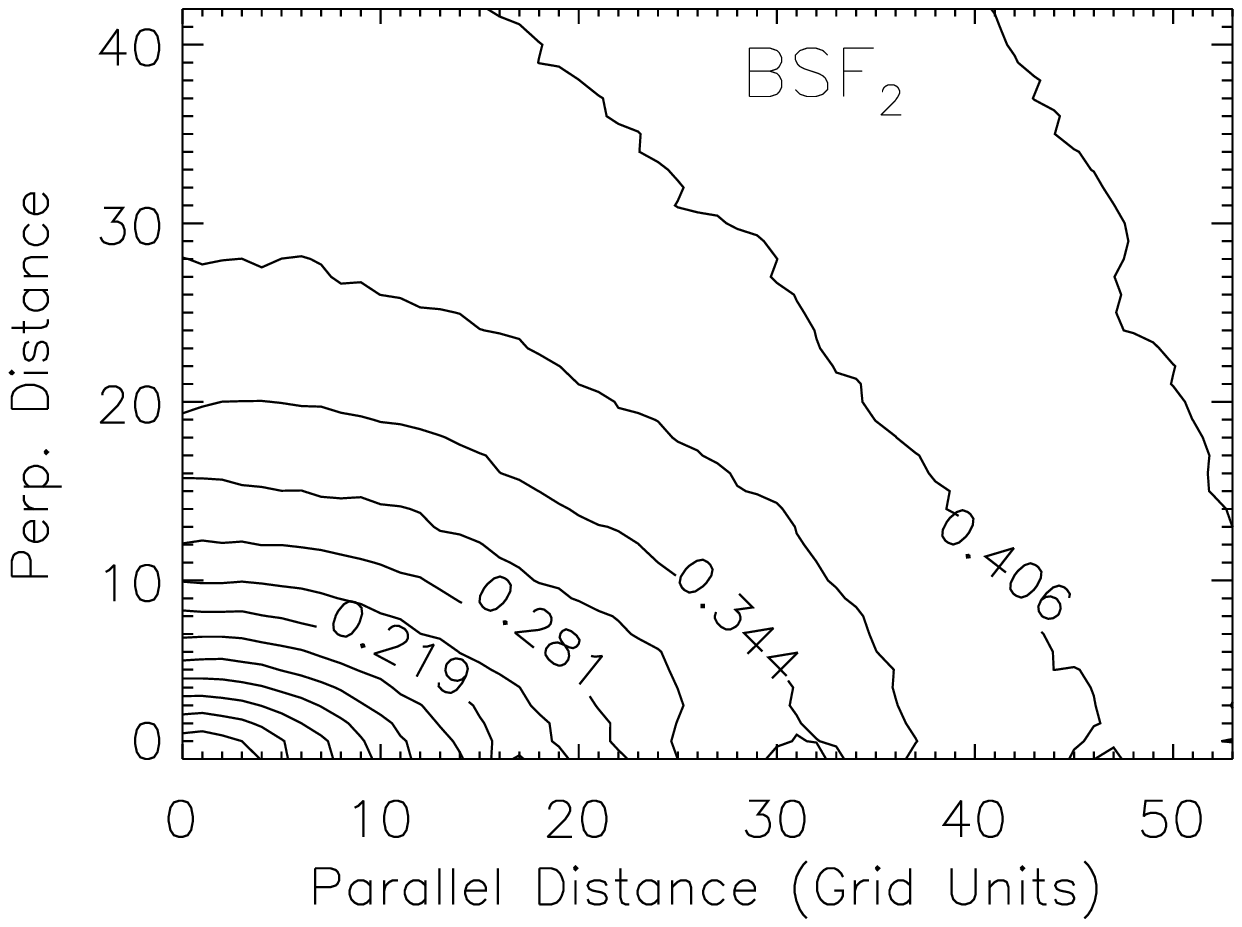}
  \caption{
      Super-Alfvenic turbulence ($M_A\sim 8$ and $M_s\sim 2.5$). 
      $V_A\equiv B_0/\sqrt{4\pi\rho}=0.1$. $a$ (sound speed) $=\sqrt{0.1}$. 
      $\delta V \sim 0.8$.
    (a){\it left:} Spectra.
    (b){\it middle:} VSF$_2$. 
    (c){\it right:} BSF$_2$. No mode decomposition is used for (b) and (c).
}
\label{fig_superAlf}
\end{figure*}
\begin{figure*}
  \includegraphics[width=0.3\textwidth]{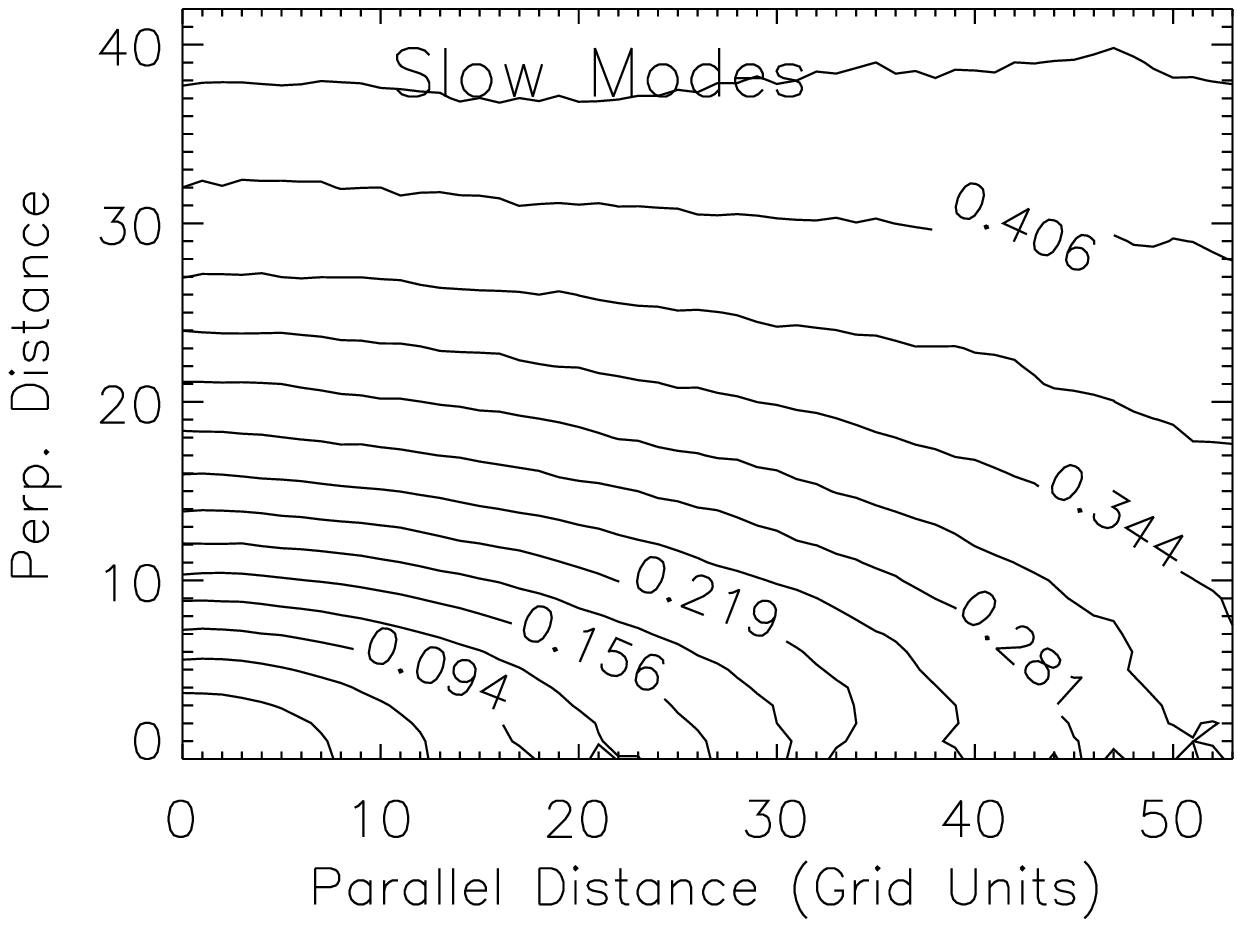}
\hfill
  \includegraphics[width=0.335\textwidth]{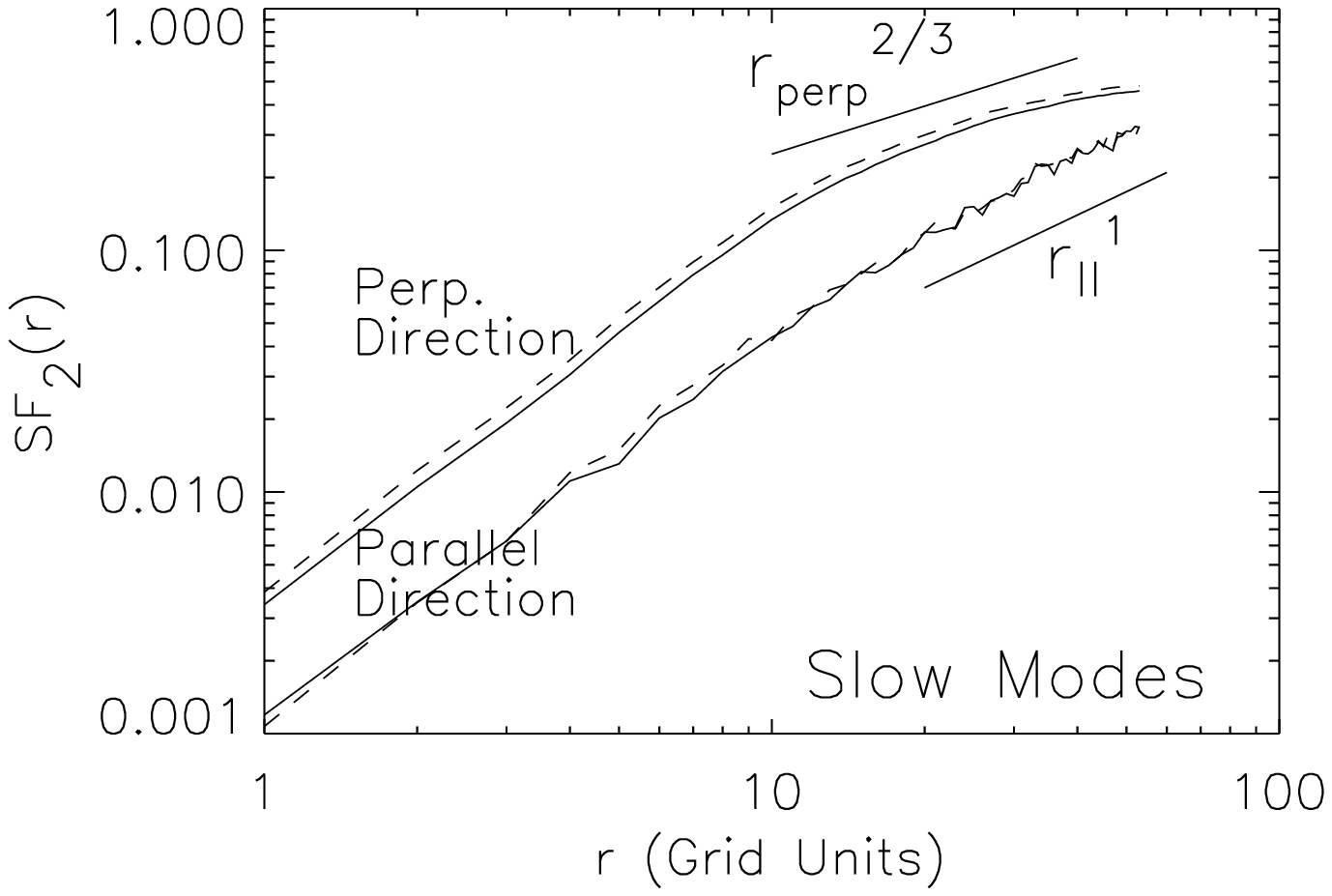}
\hfill
  \includegraphics[width=0.335\textwidth]{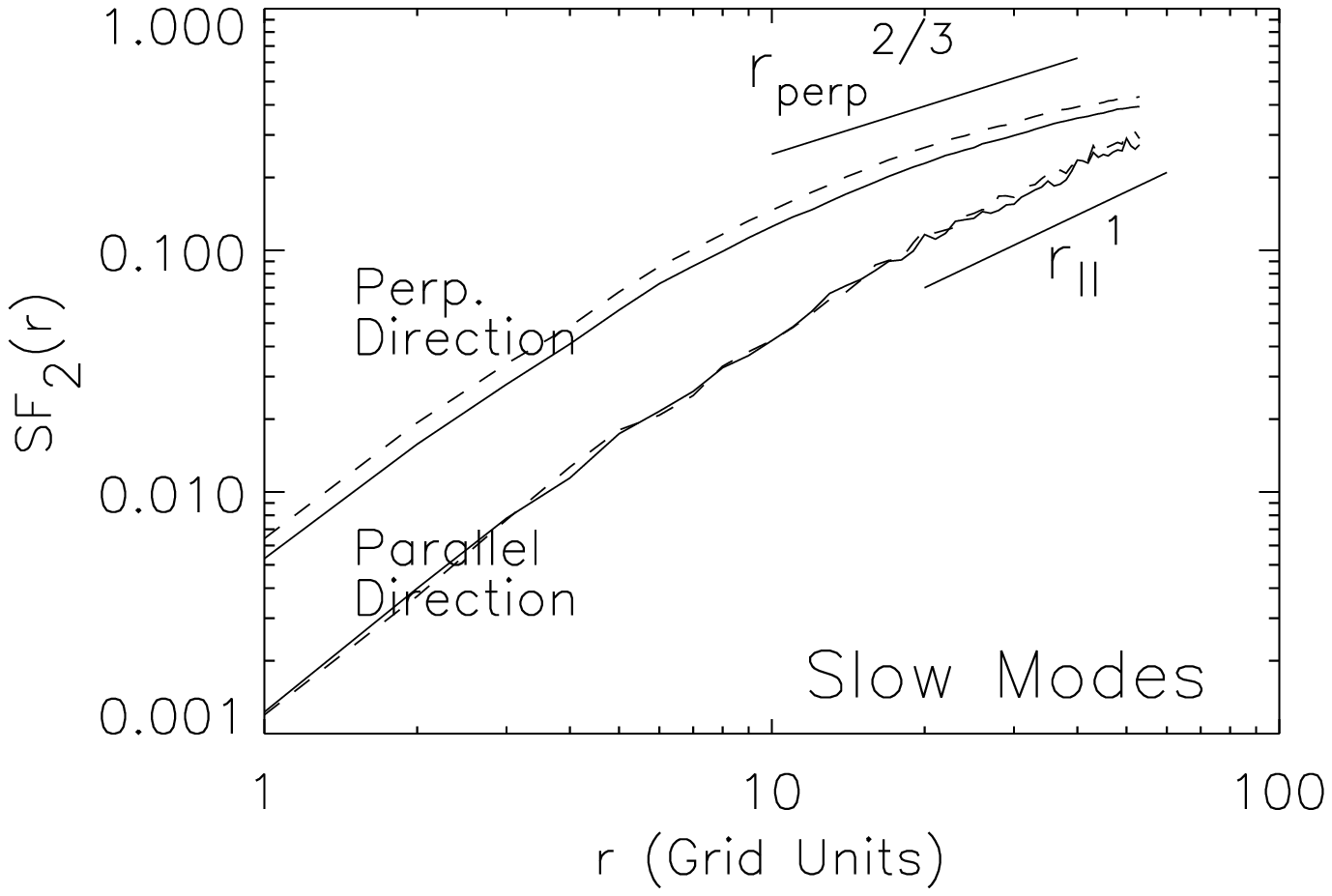}
  \caption{
      Comparison between Fourier space method and real space method.
    (a){\it left:} From real space calculation. $M_s\sim 7$.
    (b){\it middle:} Solid: Fourier space. Dashed: real space. $M_s\sim 7$.
    (c){\it right:} Similar plot for $M_s\sim$ 2.3.
}
\label{fig_comparision}
\end{figure*}

So far we considered ``sub-Alfvenic'' turbulence in which the Alfven speed 
associated with
the mean magnetic field is slightly faster than the r.m.s. fluid velocity.
If the opposite case is true (i.e. if the mean field $B_0$ is weak), 
the turbulence is 
called ``super-Alfvenic''.
        In super-Alfvenic turbulence, large scale magnetic field lines 
        can show very chaotic structures.
Whether or not the ISM turbulence is sub-Alfvenic is still
a controversial issue.

We mentioned in \S\ref{section_theoretical} that,
even in the case of super-Alfvenic turbulence,
we can find
some scale $l'$ in the turbulent cascade 
where $v_{l'}\sim B/\sqrt{4 \pi \rho}$ and we
can apply our model of sub-Alfvenic turbulence
for all smaller scales. 
Fig.~\ref{fig_superAlf} supports this idea.
The contours in the figure are the second order structure functions
(SF$_2$)
of velocity and magnetic field. We do not use mode decomposition.
Nevertheless, the velocity SF$_2$ reflects scalings of
Alfven and slow modes, because fast modes  are weaker than Alfven and 
slow modes.
As expected, anisotropy emerges at small scales.
This is very similar to incompressible case (Cho \& Vishniac 2000a).

        Fig.~\ref{fig_superAlf}(a) shows power spectra.
        We notice that the power spectra of  velocity and  magnetic field
        have different shapes.
        The velocity power spectrum is larger than  the magnetic one
        near the energy injection scale at $k\sim 2.5$.
        However, for larger k's ($k>6$), 
        the magnetic power spectrum is larger than the velocity one.
        This behavior is well known in incompressible simulations with
        unit magnetic Prandtl number 
        (see, for example, Kida, Yanase, \& Mizushima 1991;
        Cho \& Vishniac 2000a).
        A similar behavior is also observed in earlier compressible
        simulations (see, for example, Brandenburg et al. 1996).
        Cho \& Vishniac (2000a) argued that the transition from $E_v(k)>E_b(k)$
        to $E_v(k)<E_b(k)$
        occurs at a wavelength 2 or 3 time larger 
        than that of the energy injection scale.

        A careful look at Fig.~\ref{fig_superAlf}(a) reveals that,
        for $k < 8$, the power spectrum
        of velocity declines faster than Kolmogorov as $k$ increases. 
        This is not very surprising because magnetic field has more power
        than velocity at large k's and, therefore, can affect
        the velocity power spectrum at small k's.
        Kida et al. (1991) claimed that the sum of $E_v(k)+E_b(k)$ 
        roughly follows
        Kolmogorov spectrum in their incompressible simulations.
        If this is true, then the velocity power spectrum should 
        have a spectrum
        steeper than Kolmogorov at small k's because
        many simulations have shown that the magnetic power spectrum 
        is significantly
        flatter than the velocity one at small k's when the mean 
        field $B_0$ is weak
        (Kida et al. 1991; Brandenburg et al. 1996; Cho \& Vishniac 2000a).
        After $k \sim 8$, it seems that the velocity power spectrum gets
        flatter.

        Boldyrev, Nordlund, \& Padoan (2002a) also obtained velocity 
        power spectrum
        steeper than Kolmogorov in their supersonic super-Alfvenic MHD
        simulations.
        They attributed the steep spectrum to different 
        intermittency properties compared to the incompressible case.
        Since they used a different simulation set-up, we do not directly
        compare our results and theirs.
        For example, their turbulence is driven at larger scales than ours
        and their sonic Mach number is larger than ours.
        Further parameter study is absolutely necessary.

\subsection{How good is our technique?}

The technique
 described in \S\ref{section_decomposition} is statistical 
in nature.
That is, we separate each MHD mode with respect to the {\it mean} magnetic field
${\bf B}_0$.
This procedure is affected
 by the wandering of large scale magnetic field lines,
as well as density inhomogeneities\footnote{One way to 
      remove the effect by the wandering of field lines is to
      drive turbulence anisotropically in such a way as 
      $k_{\perp, L} \delta V \sim k_{\|,L} V_A$, where
      $k_{\perp,L}$ and $k_{\|,L}$ stand for the wavelengths 
      of the driving scale and
      $\delta V$ is the r.m.s. velocity.
      By increasing the $k_{\perp,L}/k_{\|,L}$ ratio,
   we can reduce the degree of mixing of different wave modes.}.

Nevertheless, we can show that our technique gives 
statistically correct results.
In low $\beta$ regime,
the velocity of a slow mode 
is nearly parallel to the {\it local} mean magnetic
field.
Therefore, for low $\beta$ plasmas, we can obtain velocity statistics
for slow modes in real space as follows.
First, we identify the direction of the {\it local} mean magnetic field
using the method described in Cho \& Vishniac (2000b).
Second, we calculate the second order structure function for slow modes
by the formula 
vSF$_2({\bf r})=< |\left( {\bf v}({\bf x}+{\bf r})-{\bf v}({\bf x}) \right)
        \cdot \hat{\bf B}_l  |^2 >$, 
where $\hat{\bf B}_l$ is the unit vector
along the {\it local} mean field.

Fig. \ref{fig_comparision}(a) shows the contours obtained by the method
for the high Mach number run.
In Fig. \ref{fig_comparision}(b), we compare the result obtained this way 
(dashed lines) and our technique 
described in \S\ref{section_decomposition} (solid lines).
We also show a similar plot for the mildly supersonic case in 
Fig. \ref{fig_comparision}(c).
These results confirm that the method described in \S\ref{section_decomposition}
gives statistically correct scaling relations.

\section{Linear Estimates of Density and Magnetic Field Fluctuations}
\begin{figure*}
  \includegraphics[width=0.33\textwidth]{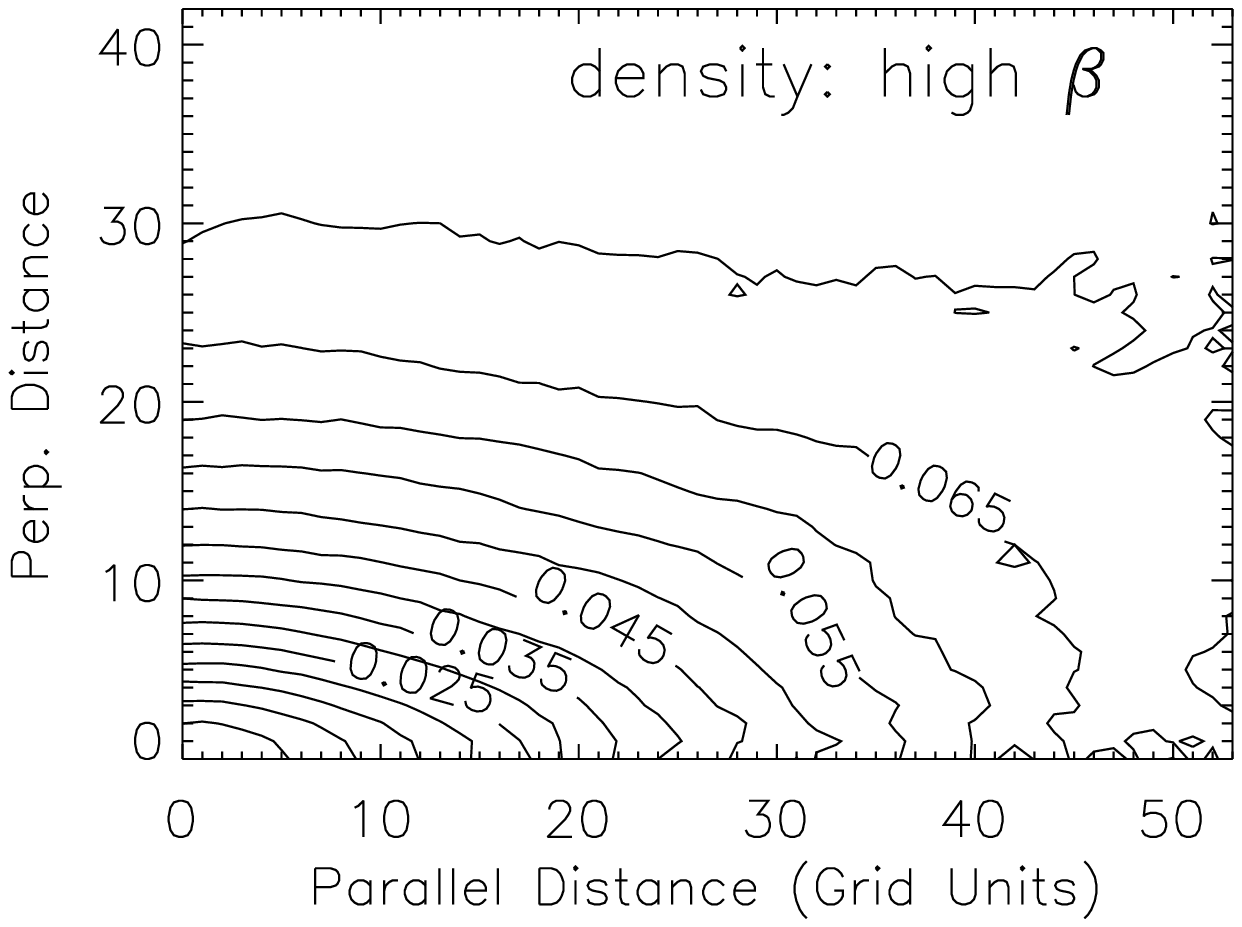}
\hfill
  \includegraphics[width=0.33\textwidth]{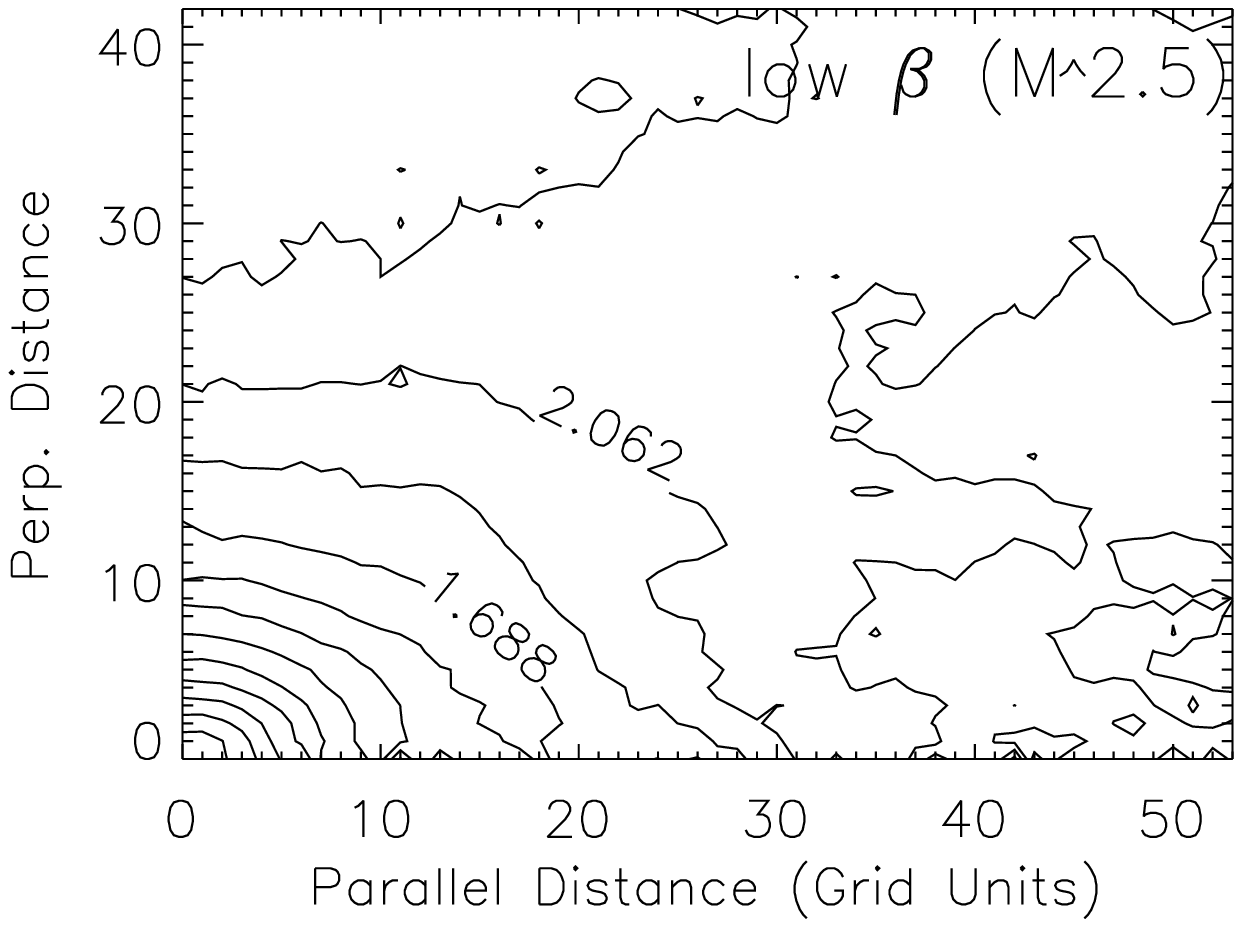}
\hfill
  \includegraphics[width=0.33\textwidth]{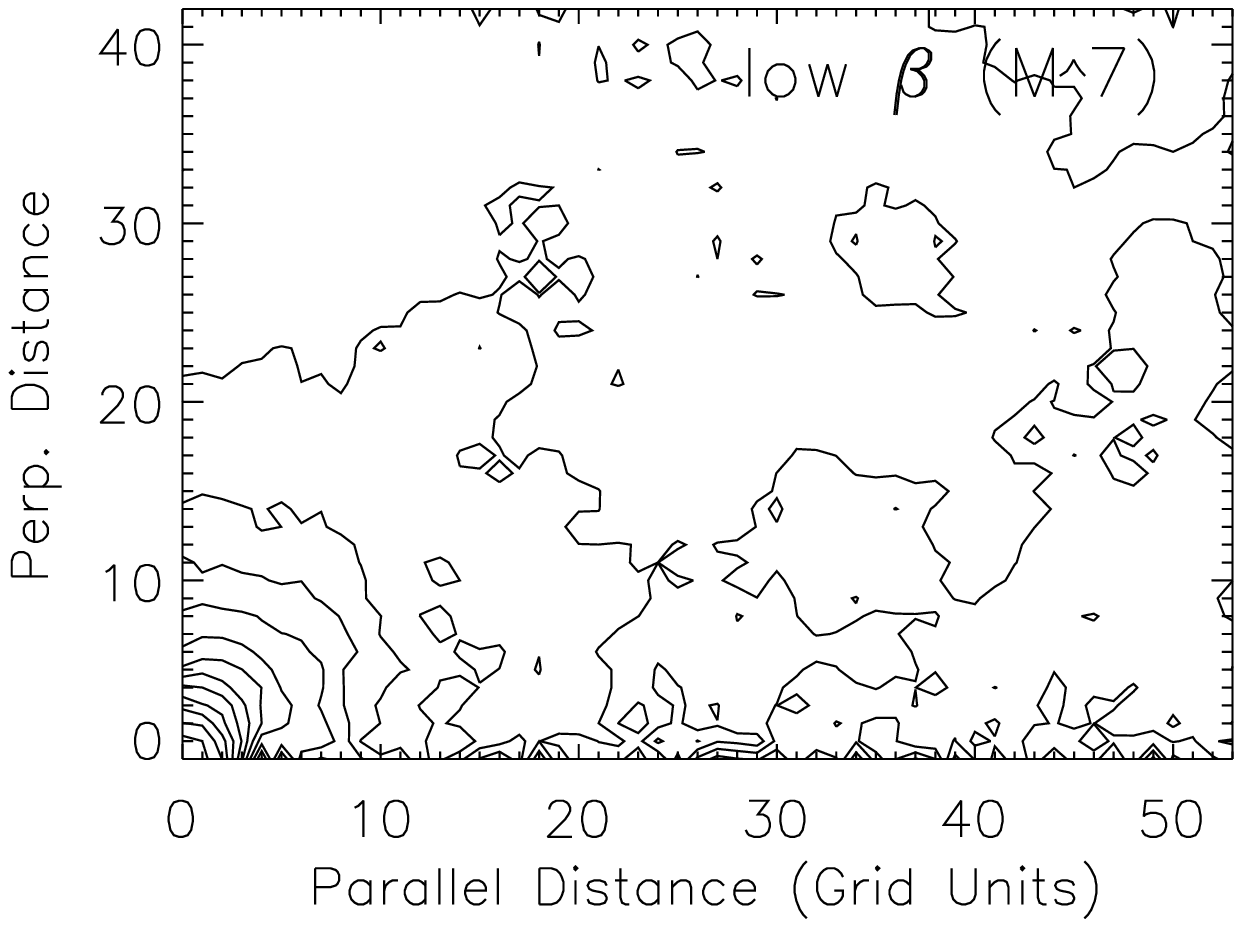}
  \caption{
       Density structures.
    (a){\it left:} $M_s\sim 0.35$ (high $\beta$).
    (b){\it middle:} $M_s\sim 2.3$ (mildly supersonic low $\beta$).
    (c){\it right:} $M_s\sim 7$ (highly supersonic low $\beta$).
}
\label{fig_density}
\end{figure*}
In this section, we estimate the magnitude of the r.m.s. fluctuations
of magnetic field and density.
Figures \ref{fig_M2}, \ref{fig_HB}, and \ref{fig_M10} show that
magnetic field and density have spectral indexes similar to those of 
velocity. We also expect that isotropy/anisotropy of magnetic field is 
similar to
that of velocity.
Therefore, we do not discuss these quantities here.
However, anisotropy of density shows different behavior.
Fig. \ref{fig_density} shows structure of density.
Density shows anisotropy for the high $\beta$ case.
But, for low $\beta$ cases, density shows more or less isotropic
structures.
We suspect that
shock formation is responsible for the isotropization of density.

To estimate the r.m.s. fluctuations, we use
the following linearized continuity and induction equations:
\begin{eqnarray}
 |\rho_k|  &=& (\rho_0 v_{k}/c) |\hat{\bf k}\cdot \hat{\bf \xi}|, \\
   |b_k|   &=& (B_0 v_{k}/c) |\hat{\bf B}_0\times \hat{\bf \xi}|,
\end{eqnarray}
where $c$ denotes propagation speed of slow or fast wave (equation (\ref{c_sf})).
{}From this, we obtain the r.m.s. fluctuations
\begin{eqnarray}
 (\delta \rho/\rho_0)_s  
       &=&  (\delta V)_s
           \langle |\hat{\bf k}\cdot \hat{\bf \xi}_s/c_s | \rangle, \\
 (\delta \rho/\rho_0)_f  
       &=& (\delta V)_f
           \langle |\hat{\bf k}\cdot \hat{\bf \xi}_f/c_f| \rangle, \\
 (\delta B/B_0)_s  
       &=& (\delta V)_s 
           \langle |\hat{\bf B}_0\times \hat{\bf \xi}_s/c_s| \rangle, \\
 (\delta B/B_0)_f  
       &=& (\delta V)_f
           \langle |\hat{\bf B}_0\times \hat{\bf \xi}_f/c_f| \rangle, 
\end{eqnarray}
where angled brackets denote a proper Fourier space average.
Generation of slow and fast modes velocity ($(\delta V)_s$ and $(\delta V)_f$)
depends on driving force.
Therefore, we may simply assume that 
\begin{equation}
  (\delta V)_{A} \sim (\delta V)_{s} \sim (\delta V)_{f},  \label{vavsvf}
\end{equation}
where we ignore constants of order unity.
However, when we consider mostly incompressible driving,
the generation of fast modes follows equation (\ref{eq_high2}).
In this case, the amplitude of fast mode velocity
is reduced by a factor of 
$ \left[ \frac{ V_A^2 + a^2 }{ (\delta V)^2_A } 
        \frac{ (\delta V)_A }{ V_A }   \right]^{-1/2}$:
\begin{equation}
  (\delta V)_{A} \sim (\delta V)_{s} 
   \sim  
   \left[ \frac{ V_A^2 + a^2 }{ (\delta V)^2_A } 
        \frac{ (\delta V)_A }{ V_A }   \right]^{1/2}
 (\delta V)_{f}.
   \label{vavsvf2}
\end{equation}
When we assume $(\delta V)_{A}\sim V_A$,
equation (\ref{vavsvf2}) reduces to equation (\ref{vavsvf})
in low $\beta$ plasmas.

\subsection{Low-$\beta$ case}
In this limit, $c_s \sim a \cos\theta$ and $c_f \sim V_A$.
Using equations (\ref{xis_lowbeta}) and (\ref{xif_lowbeta}),
we obtain
\begin{eqnarray}
 (\delta \rho/\rho_0)_s  
       &\sim& (\delta V)_s 
           \langle |\cos\theta/c_s| \rangle \sim (\delta V)_s/a, \\
 (\delta \rho/\rho_0)_f  
       &=& (\delta V)_f 
           \langle |\sin\theta/c_f| \rangle \sim (\delta V)_f/V_A, \\
 (\delta B/B_0)_s  
       &=& (\delta V)_s 
           \langle |\alpha \cos\theta \sin\theta/c_s| \rangle 
            \sim \alpha (\delta V)_s/a, \\
 (\delta B/B_0)_f  
       &=& (\delta V)_f 
           \langle | 1/c_f| \rangle \sim (\delta V)_f/V_A,   \label{bf_lowbeta}
\end{eqnarray}
where we ignore $\cos\theta$'s or $\sin\theta$'s.

When we assume 
$(\delta V)_{A}\sim (\delta V)_{s}\sim (\delta V)_{f}\sim V_A$, we get
\begin{eqnarray}
 (\delta \rho/\rho_0)_s  &\sim& M_s, \\
 (\delta \rho/\rho_0)_f  &=& \sqrt{\beta} M_s, \\
 (\delta B/B_0)_s  &=& \beta M_s, \\
 (\delta B/B_0)_f  &=& \sqrt{\beta} M_s
\end{eqnarray}
Therefore, in low $\beta$ plasmas, slow modes give rise to most of 
density fluctuations
(CL02).
On the other hand, magnetic fluctuation by slow modes is smaller than that
by fast modes by a factor of $\sqrt{\beta}$.

\subsection{High-$\beta$ case}
In this limit, $c_s \sim V_A \cos\theta$ and $c_f \sim a$.
Using equations (\ref{xis_highbeta}) and (\ref{xif_highbeta}),
we obtain
\begin{eqnarray}
 (\delta \rho/\rho_0)_s  
       &\sim& (\delta V)_s 
           \langle |\cos\theta \sin\theta/(\alpha c_s)| \rangle \nonumber \\
           &\sim& (V_A/a)(\delta V)_s/a, \\
 (\delta \rho/\rho_0)_f  
       &=& (\delta V)_f 
           \langle |1/c_f| \rangle \sim (\delta V)_f/a, \\
 (\delta B/B_0)_s  
       &=& (\delta V)_s 
           \langle |\cos\theta/c_s| \rangle 
            \sim  (\delta V)_s/V_A, \\
 (\delta B/B_0)_f  
       &=& (\delta V)_f 
           \langle | \sin\theta/c_f| \rangle \sim (\delta V)_f/a,  
\end{eqnarray}
where we ignore $\cos\theta$'s or $\sin\theta$'s.

Let us just assume that 
$(\delta V)_{A}\sim (\delta V)_{s}\sim V_A  \sim M_s^{-1} (\delta V)_{f}$
(cf. equation (\ref{vavsvf2})).
Then we have
\begin{eqnarray}
 (\delta \rho/\rho_0)_s  &\sim&  M_s/\sqrt{\beta} \sim M_s^2, \\
 (\delta \rho/\rho_0)_f  &\sim&  M_s^2, \\
 (\delta B/B_0)_s         &=& O(1), \\
 (\delta B/B_0)_f         &=& M_s^2.
\end{eqnarray}
The density fluctuation associated with slow modes
is $\sim M_s^2$, when
$(\delta V)_s\sim (\delta V)_{A} \sim V_A$.
This is consistent with Zank \& Matthaeus (1993).
The ratio of $(\delta \rho)_s$ to $(\delta \rho)_f$ is of order unity.
Therefore, both slow and fast modes give rise to similar amount of density
fluctuations.
Note that this argument is of order-of-magnitude in nature.
In fact, in our simulations for the high $\beta$ case, 
the r.m.s. density fluctuation by slow modes
is about twice as large as that by fast modes.
When we use equation (\ref{vavsvf}), 
we have a different result:
$ (\delta \rho)_s \sim (V_A/a)(\delta \rho)_f  < (\delta \rho)_f $.
It is obvious that
slow modes dominate magnetic fluctuations:
$(\delta B)_s > (\delta B)_f$ for both equations 
(\ref{vavsvf}) and (\ref{vavsvf2}).

\section{Viscosity-damped Regime of MHD Turbulence} 
 \label{sect_new}

In hydrodynamic turbulence viscosity sets a minimal scale for
motion, with an exponential suppression of motion on smaller
scales.  Below the viscous cutoff the kinetic energy contained in a 
wavenumber band is 
dissipated at that scale, instead of being transferred to smaller scales.
This means the end of the hydrodynamic cascade, but in MHD turbulence
this is not the end of magnetic structure evolution.  For 
viscosity much larger than resistivity,
$\nu\gg\eta$, there will be a broad range of
scales where viscosity is important but resistivity is not.  
On these
scales magnetic field structures will be created 
by the shear from non-damped turbulent motions, which
amounts essentially to the shear from the smallest undamped scales.
The created magnetic structures would evolve through
generating small scale motions.
As a result, we expect
a power-law tail in the magnetic energy distribution, 
rather than an exponential
cutoff.  
      Cho, Lazarian, \& Vishniac (CLV02b) performed
      numerical simulations of turbulence in this regime 
      threaded by  a strong  ($B_0/\sqrt{4 \pi \rho}\sim \delta V$)
      mean magnetic field 
      and reported that
      this regime possesses completely different scaling relations
      and anisotropic structures compared with ordinary
      MHD turbulence.
Further research showed that there is a smooth connection between this
regime and small scale turbulent dynamo in high magnetic 
Prandtl number fluids
(see Schekochihin et al.~2002)\footnote{
      It is worth noting that, 
      motivated by the analogy between time evolution equations for 
      magnetic field and for vorticity,
      Batchelor (1950) first argued that
      small magnetic fields can be amplified when viscosity is larger than
      magnetic diffusion (i.e. magnetic Prandtl number $>$ 1).
      Although the analogy was later proved to be physically wrong,
      the high magnetic Prandtl number dynamo has been studied by
      many researchers
      (e.g. Kulsrud \& Anderson 1992; Kinney et al. 2000).}.

\begin{figure*}
  \includegraphics[width=0.49\textwidth]{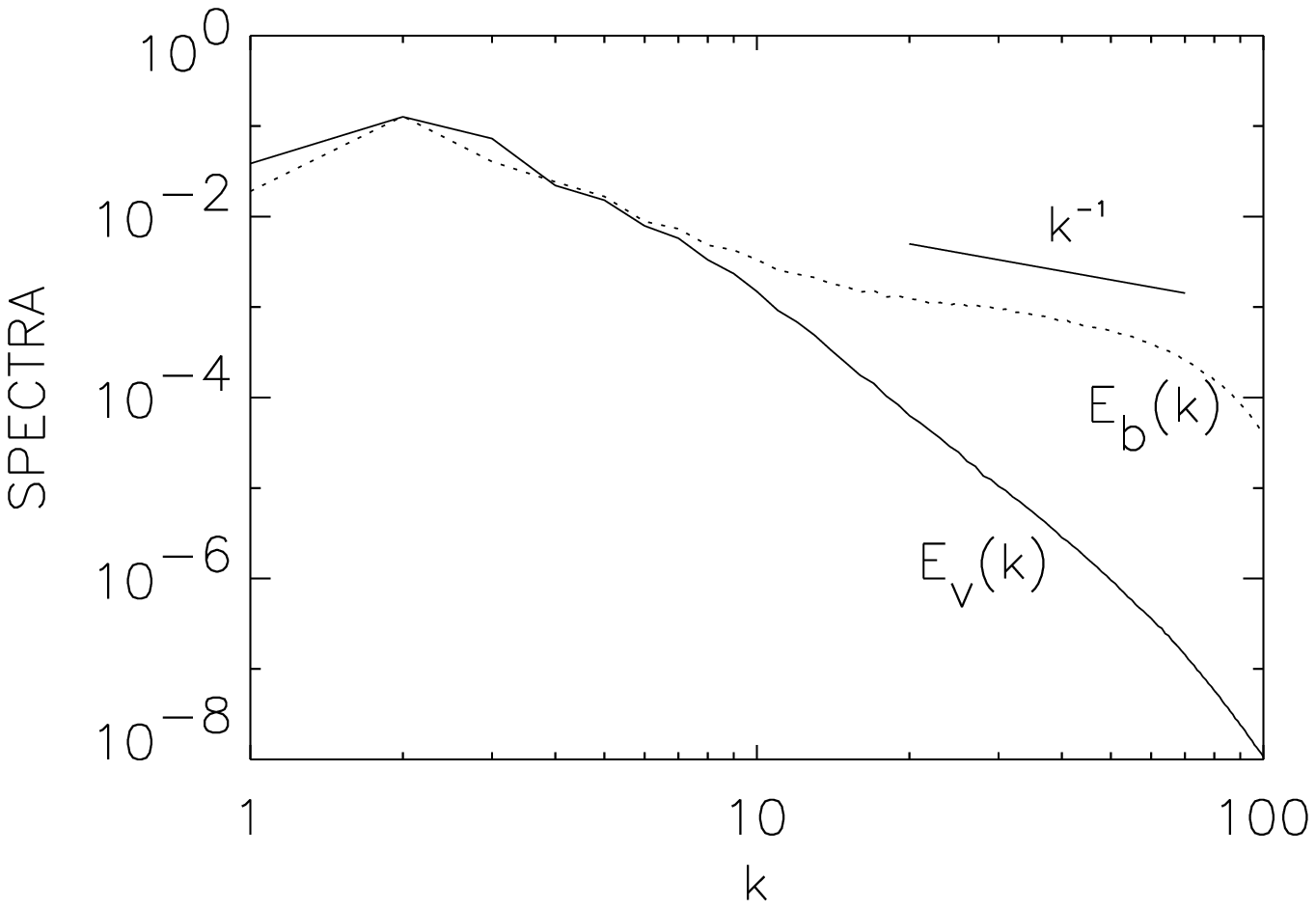}
\hfill
  \includegraphics[width=0.49\textwidth]{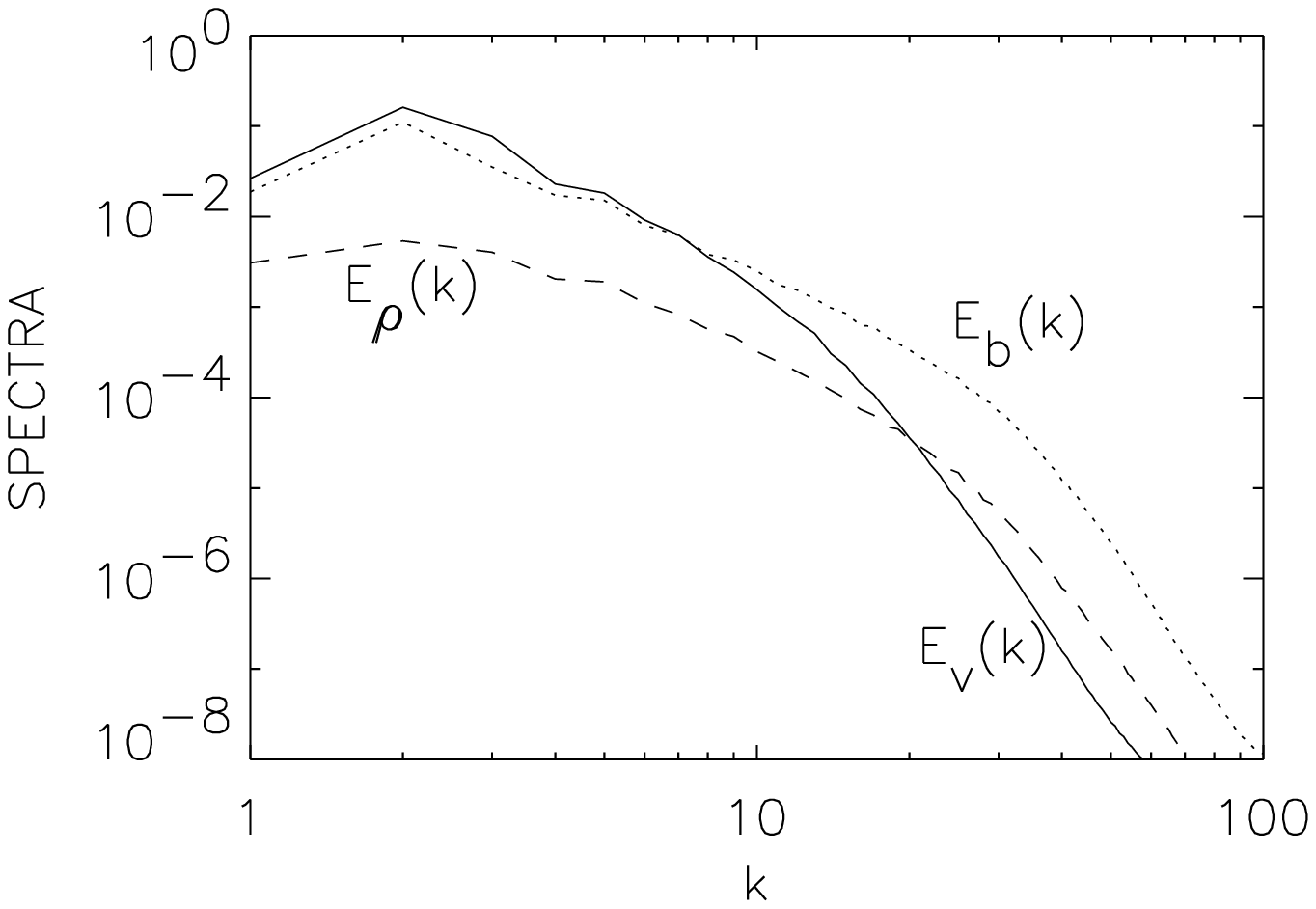}
  \caption{
      Viscous damped regime (viscosity $>$ magnetic diffusivity).
      Due to large viscosity, velocity damps after $k\sim10$.
    (a) {\it Left:} Incompressible case with $384^3$ grid points. 
        Magnetic spectra show a shallower slope ($E_b(k)\propto k^{-1}$)
        below the velocity damping scale.
        We achieve a very small magnetic diffusivity through the use
        hyper-diffusion.
       {}From Cho, Lazarian, \& Vishniac (2002b).
    (b) {\it Right:} Compressible case with $216^3$ grid points.
        Magnetic and density spectra show structures below the
        velocity damping scale at $k\sim10$.
        The structures are less obvious than the incompressible case
        because it is relatively hard to
        achieve very small magnetic diffusivity in the compressible run.
}
\label{fig_viscous}
\end{figure*}
In partially ionized gas
neutrals produce viscous damping of turbulent motions. 
In the Cold Neutral Medium (see Draine \& Lazarian 1999 for a list of
the idealized phases) this produces damping on the scale of a fraction of
a parsec. The magnetic diffusion in those circumstances is
still negligible and exerts an influence only 
at the much smaller scales, $\sim 100km$. 
Therefore, there is a large
range of scales where the physics of the turbulent cascade 
is very different from 
the conventional MHD turbulence picture.

CLV02b explored this regime numerically with a grid of $384^3$ and a 
physical viscosity for velocity damping. 
The kinetic Reynolds number was around 100. 
We achieved a very small magnetic diffusivity by the use of
hyper-diffusion.
The result is presented in  Fig.~\ref{fig_viscous}a.
A theoretical model for this regime and its
consequences for stochastic reconnection (Lazarian \& Vishniac 1999) 
can be found in Lazarian, Vishniac, \& Cho (2003). 
It explains the spectrum $E(k)\sim k^{-1}$ as a cascade of magnetic
energy to small scales under the influence of shear at the 
marginally damped scales. 
     The spectrum is similar to that of the viscous-convective range of
     passive scalar in hydrodynamic turbulence (see, for example,
     Batchelor 1959; Lesieur 1990), although the study in Lazarian,
     Vishniac, \& Cho (2003) suggests that the physical origin of it 
     is different.
     A study confirming that the $k^{-1}$ spectrum
        is not a bottleneck effect is presented in 
        Cho, Lazarian, \& Vishniac (2003b).
The mechanism is based on the solenoidal
motions and therefore the compressibility should not alter the physics
of this regime of turbulence.  

We show our results for the compressible fluid in Fig~\ref{fig_viscous}b. 
We use the same physical viscosity as in incompressible case (see
CLV02b).
We rely on numerical diffusion, which is much smaller than physical
viscosity, for magnetic field.
The inertial 
range is much smaller due to numerical reasons, but it is clear that
the viscosity-damped regime of MHD turbulence persists. 
The magnetic fluctuations,
however, compress the gas and thus cause fluctuations in density.
This is a new (although expected) phenomenon compared to our earlier
incompressible calculations. These density fluctuations may have important
consequences for the small scale structure of the ISM.

\section{Astrophysical Implications of Our Results}

\subsection{Parameter range explored}

Parameters of astrophysical plasmas differ substantially for different
astrophysical systems:
from extremely high $\beta$ to extremely low $\beta$. Turbulence 
in some systems is expected to be superAlfvenic, in others it is expected
to be subAlfvenic. Moreover, there is an ongoing controversy on what to
expect and where. For instance, while high $\beta$ was considered a
default for many phases of Milky Way ISM, recent observations by
Beck (2002) suggest that the plasmas there may be low $\beta$. Therefore
it is essential to have clear understanding of MHD turbulence for as 
large parameter space as possible.

{\bf SuperAlfvenic regime.---}
We have argued above that the difference between superAlfvenic and 
subAlfvenic
turbulence is not as substantial as it is frequently thought. 
The difference between the two regimes stems from
ratio of magnetic field to kinetic energies. However, 
as we mentioned in \S2, if kinetic energy density exceed magnetic
field energy density, the hydromagnetic motions drive turbulent
dynamo. This enhances magnetic field energy density up to 
approximately equipartition value. Thus the difference between the
superAlfvenic and subAlfvenic regimes amounts not to the energy of
the magnetic field, but to the global level of field organization,
e.g. to the magnetic field reversals etc. Our results in \S\ref{sect_super} 
suggest that the
basic properties of the MHD turbulence in the subAlfvenic
and superAlfvenic regimes are similar.
This, however, does not preclude the intermittency of MHD turbulence
being very different. The latter property can be tested using scalings 
of the higher order  velocity correlations 
$SF_p({\bf r}) \equiv \langle 
  |{\bf v}({\bf x}+{\bf r})-{\bf v}({\bf x}) |^p\rangle \propto r^{\zeta (p)}$. 
The 
corresponding scaling suggested by She-Leveque (1994) contains
three parameters (see Politano \& Pouquet 1995; M\"{u}ller \& Biskamp 2000): 
$g$ is related to the scaling $\delta v_l\sim l^{1/g}$, $x$ related to
the energy cascade rate $t_l^{-1}\sim l^{-x}$, and C, the co-dimension of the
dissipative structures:
\begin{equation}
\zeta(p)=p/g(1-x)+C\left( 1-(1-x/C)^{p/g} \right).
\label{She-Leveque}
\end{equation}
M\"{u}ller \& Biskamp (2000) proposed that 3D 
{\it incompressible} MHD turbulence for {\it zero mean field}
 has Kolmogorov $g$ and $x$, while the
the dissipation happens in the sheet-like structures, i.e. $C=3-2=1$. Using
eq.~(\ref{She-Leveque}) they obtained an excellent fit for their 
{\it incompressible} data. Later Boldyrev (2002) made the same 
assumption\footnote{The physical motivation of M\"{u}ller \& Biskamp (2000)
and Boldyrev (2002) for choosing the same value of $C$ are different,
however. Boldyrev (2002) identifies the dissipation
structures with shocks, which are absent in {\it incompressible}
simulations by M\"{u}ller \& Biskamp (2000).} 
that $C=1$ (and same $g$ and $x$) for
the compressible turbulence and Boldyrev, Nordlund, \& Padoan (2002a)
obtained
an excellent fit to their
{\it compressible} data. 
     It appears surprising that incompressible MHD 
     (M\"{u}ller \& Biskamp 2000) and
     supersonic compressible MHD (Boldyrev et al. 2002a) have similar
     intermittency structures.
     This issue is discussed in
     Cho, Lazarian, \& Vishniac (2003b).

We would like to note, however, that our considerations about superAlfvenic
turbulence are not applicable to the transient regimes when superAlfvenic
motions are in the process of generating magnetic field and magnetic
energy does not have time to come to a rough equipartition to the kinetic
energy. Some parts of the ISM can well be in the transient regime for
which temporary $\rho V^2/2\gg B^2/8\pi$.  

\begin{figure*}
  \includegraphics[width=0.24\textwidth]{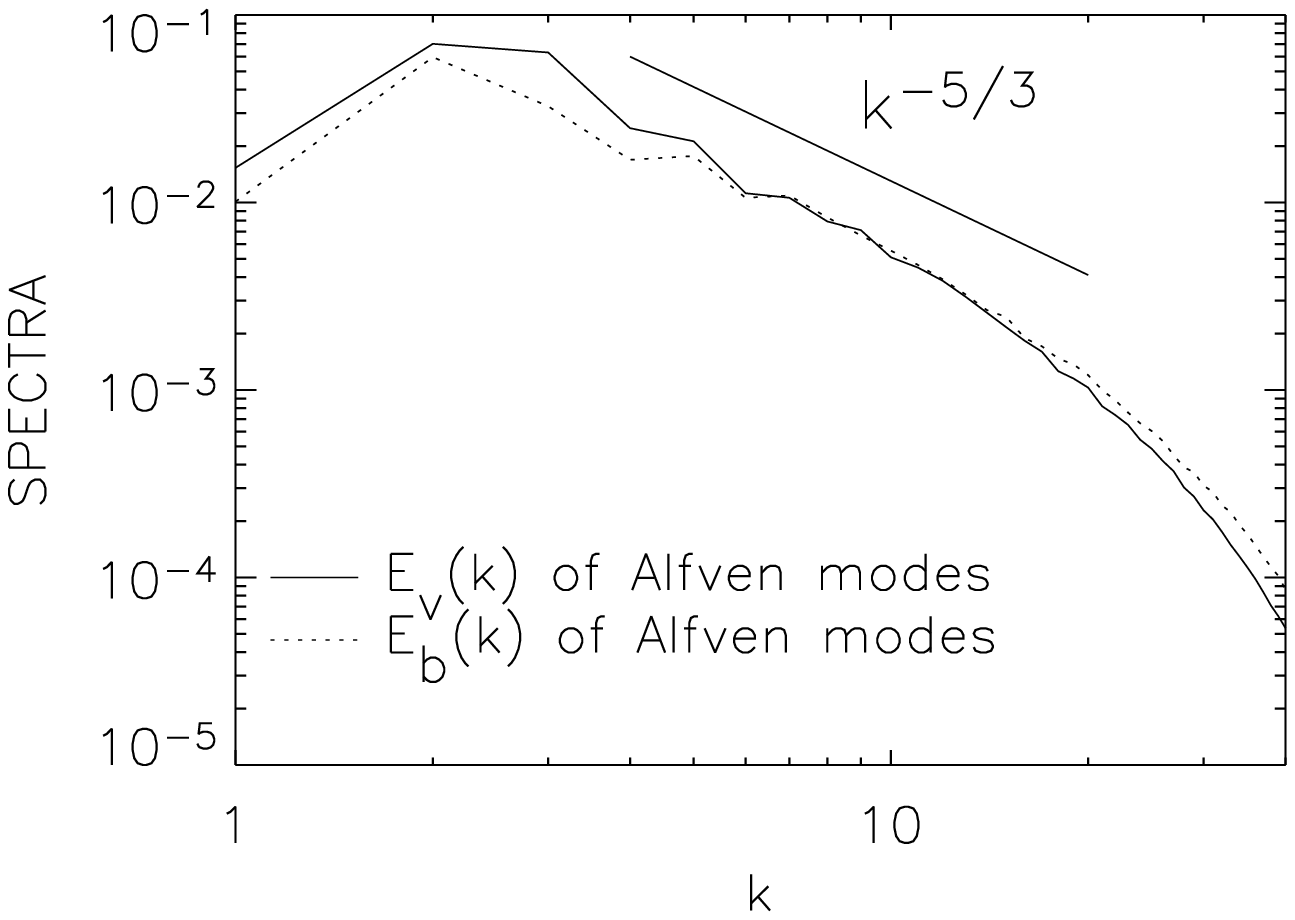}
\hfill
  \includegraphics[width=0.24\textwidth]{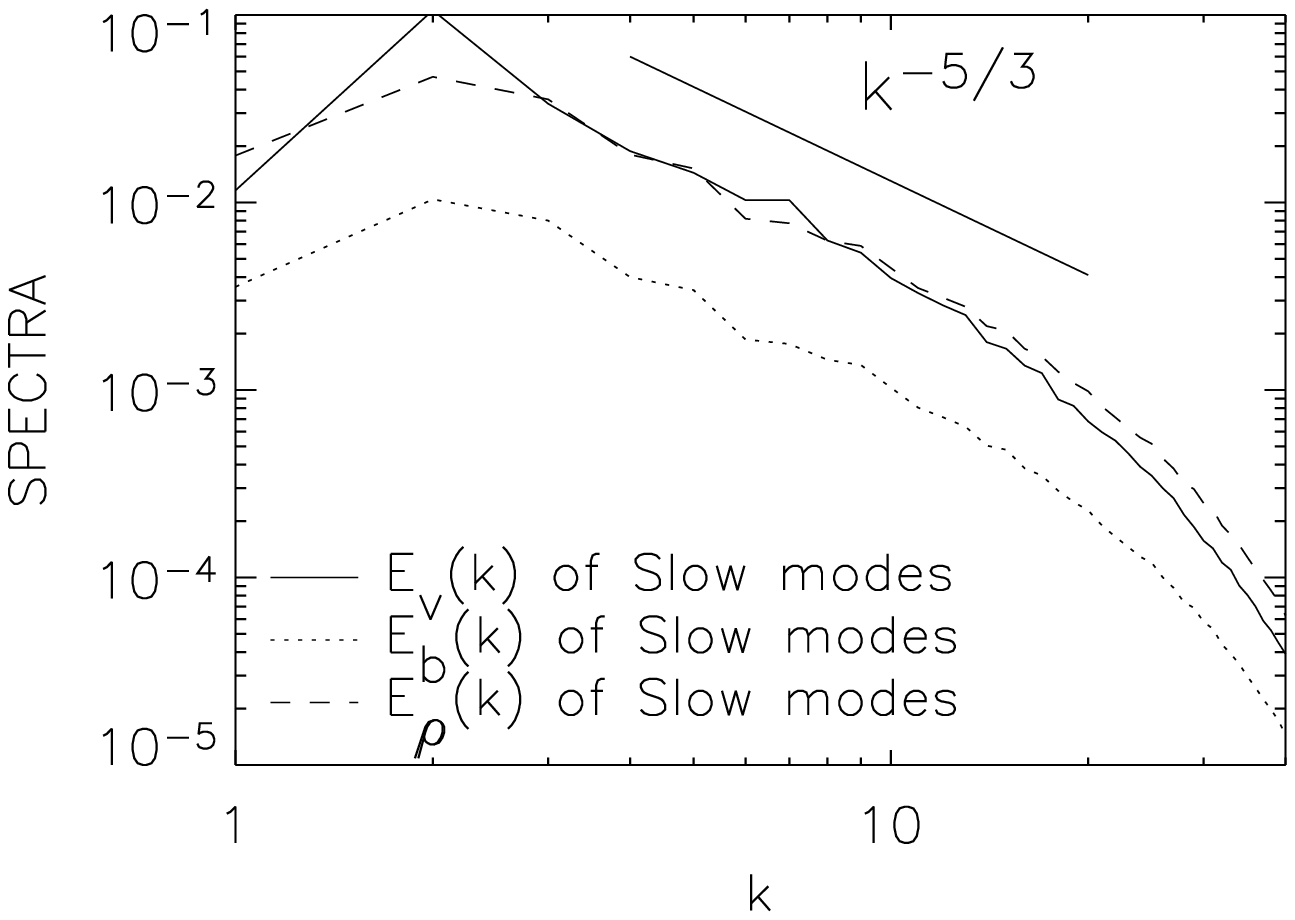}
\hfill
  \includegraphics[width=0.24\textwidth]{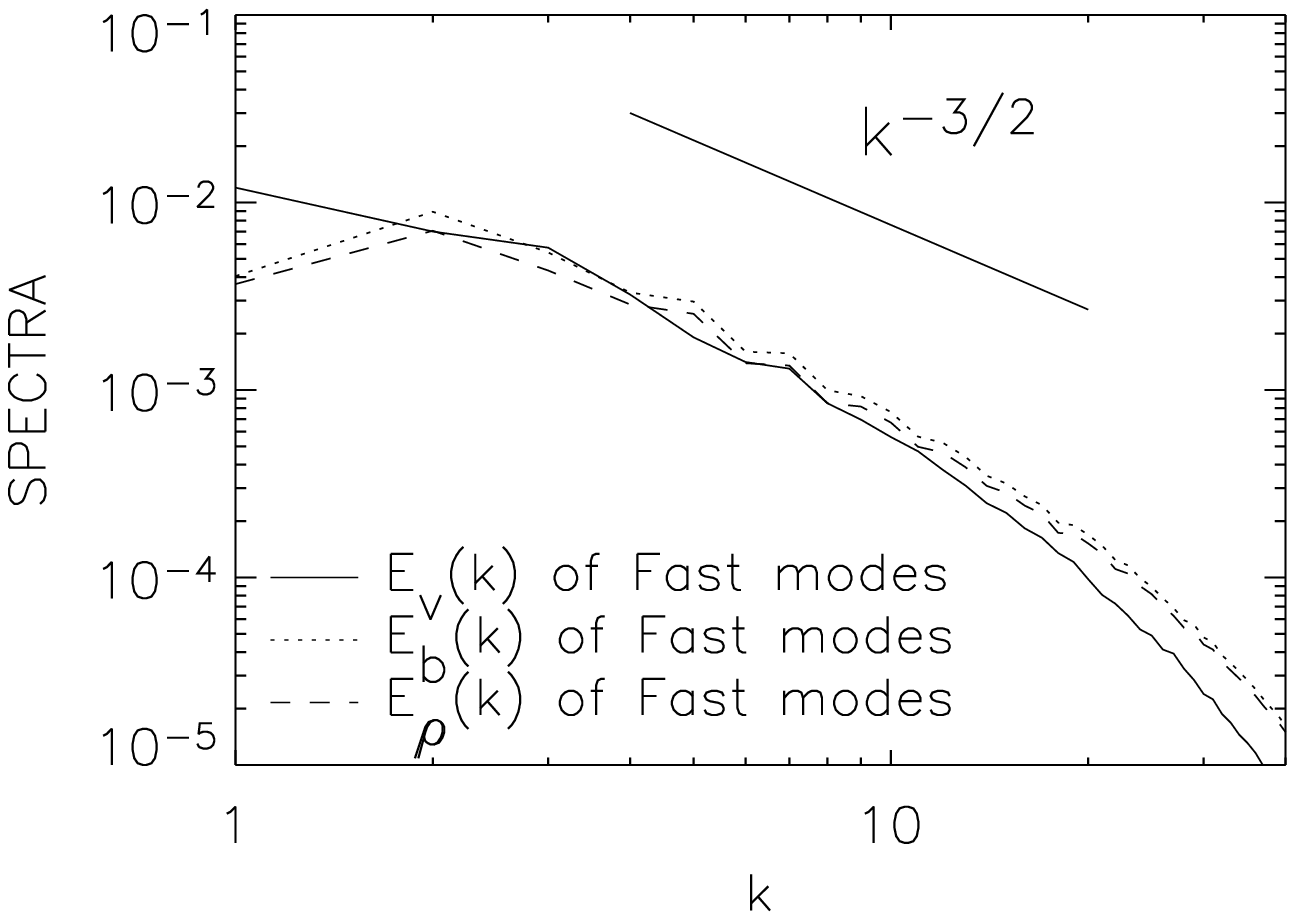}
\hfill
  \includegraphics[width=0.24\textwidth]{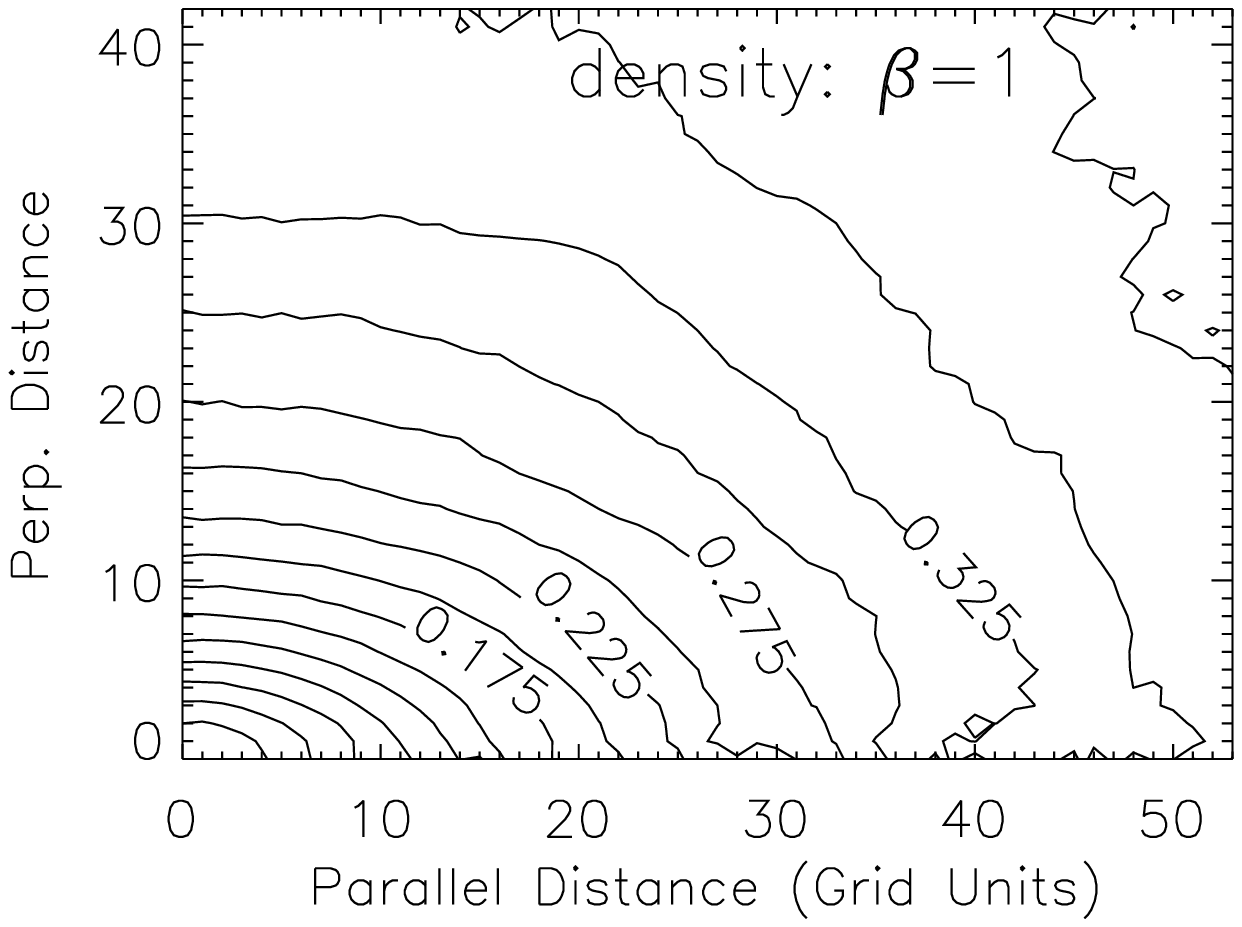}
  \caption{
      $\beta=1$ case.
    (a){\it left:} Alfven mode spectra. 
    (b){\it second-left:} Slow mode spectra.
    (c){\it second-right:} Fast mode spectra.
    (d){\it right:} Density structure.
}
\label{fig_beta1}
\end{figure*}

{\bf $\beta=1$ case.---}
We provided theoretical considerations for both $\beta \gg 1$ and
$\beta \ll 1$ cases. What about the intermediate cases of $\beta\sim 1$?
Would the scaling of modes and mode coupling be different?
To test this case we performed calculations for $\beta=1$.
The results in Fig.~\ref{fig_beta1} show that
the scaling relations that hold true for mildly low $\beta$ regime
are also applicable for $\beta=1$ regime.

\subsection{Fundamental questions}

{\bf How fast does MHD turbulence decay?} 
This question has fundamental
implications for {\it star formation} (see McKee 1999). 
Indeed, it was thought originally that
magnetic fields would prevent turbulence from fast decay. Later 
(see Mac Low et al.~1998;
Stone et al.~1998; and review by Vazquez-Semadeni et al.~2000) 
this was reported not to be true. However, fast decay was erroneously
associated with the coupling between compressible and incompressible 
modes. The idea was that incompressible motions 
quickly transfer their energy to the compressible modes, which get
damped fast by direct dissipation (presumably through shock formation).

Our calculations support the conjecture given by eq.~(\ref{eq_high2}).
According to it the coupling of Alfven and compressible motions is important
only at the energy injection scales where $\delta V_l\sim V_A$. 
As the turbulence evolves the perturbations become smaller and the
coupling less efficient. Typically for numerical simulations
the inertial range is rather small and this could explain why
marginal coupling of modes was not noticed.

Our results show that MHD turbulence damping does not depend on whether
the fluid is compressible or not. The incompressible motions damp also within
one eddy turnover time\footnote{
        It is generally believed that decay of turbulence energy 
        follows a power-law.
        A possible expression is
        $E(t) \sim E_0 / [ 1 + C(t-t_0)]^{n}$ (C. McKee, private communication),
        where $E_0$ is the energy at the initial time $t_0$.
        Our claim is that decay of turbulence follows
        $E(t) \sim E_0 / [ 1 + C^{\prime}(t-t_0)/t_{cas,0}]^{n}$, where
        $t_{cas,0}$ is the eddy turn-over time at $t=t_0$ and
        $C^{\prime}$ is a constant of order unity.
}.
This is the consequence of the fact that within
the strong turbulence\footnote{
    For a formal definition of
    strong, weak and intermediate turbulence, see Goldreich \& Sridhar (1997)
    and CLV03a, but
    here we just mention in passing that in most astrophysically important
    cases the MHD turbulence is ``strong''.} 
mixing motions perpendicular
to magnetic field are hydrodynamic to high order (CLV02a) and the 
cascade of energy induced by those motions is similar to the hydrodynamic 
one, i.e. energy cascade happens within an eddy turnover time.      

The reported (see Mac Low et al.~1998) decay of the {\it total}
energy of turbulent motions $E_{tot}$ follows $t^{-1}$ which can
be understood if we account for the fact that the energy is being
injected at the scale smaller than the scale of the system. Therefore
some energy originally diffuses to larger scales through the inverse 
cascade. Our calculations stimulated by illuminating 
discussions with Chris McKee show that if this energy 
transfer is artificially
prevented by injecting the energy on the scale of the computational box, the
scaling of $E_{tot}$  becomes closer to $t^{-2}$.

{\bf Can compressible MHD turbulence decay slowly?} 
Incompressible MHD computations
(see Maron \& Goldreich 2001; CLV02a) show that the
rate of turbulence decay depends on the degree of turbulence 
imbalance\footnote{This quantity is also called cross helicity
(see Matthaeus, Goldstein, \& Montgomery 1983).}, i.e.
the difference in the energy fluxes moving in opposite directions.  
The strongly imbalanced incompressible turbulence was shown to persist
longer than its balanced counterpart. This enabled CLV02a to speculate that 
this may enable energy transfer between clouds and may
explain the observed turbulent linewidths of GMCs without evident 
star formation. Imbalanced turbulence can also make it possible to 
transfer energy from the galactic disk to heat the Reynolds layer 
(see Reynolds 1988).

Our results above show a marginal coupling of compressible and 
incompressible
modes. This is suggestive that the results obtained in incompressible 
simulations are applicable to compressible environments
if amplitudes of perturbations are not large.
The complication arises from the existence of the parametric instability 
(Del Zanna, Velli, \& Londrillo 2001) that happens as the density 
perturbations reflect Alfven waves
and grow in amplitude. This instability eventually controls the degree
of imbalance that is achievable. However, the growth rate of the instability
is substantially slower than the Alfven wave oscillation rate. Therefore,
if we take into account that interstellar sources are intermittent not
only in space, but also in time, the transport of turbulent 
energy described in CLV02a seems feasible. Here we mention that the
growth of the parametric instability described above may provide
an alternative explanation for the observed infall motions of the cloud cores
(Tafalla et al.~1998). Earlier those motions were explained as
arising from linear damping of Alfven waves (Myers \& Lazarian 1998).

{\bf Is the correlation between magnetic field and density tight in
MHD turbulence?}
In the traditional static paradigm of the ISM, density and magnetic
field increase simultaneously as clouds contract. 
Introduction of turbulence in the picture
of ISM complicates the analysis (see discussion in Vazquez-Semadeni et al.
2000; CLV03a). Our results confirm earlier claims
(e.g. Passot  \& Vazquez-Semadeni 2003)  that magnetic field
- density correlations may be weak. First of all, some magnetic field
fluctuations are related to Alfvenic turbulence which does not compress
the medium. Second, slow modes in low $\beta$ plasmas are essentially
density perturbations that propagate along magnetic field and which
marginally perturb magnetic fields. Moreover, our calculations 
(see Fig.~\ref{fig_density})
show that at substantially high Mach numbers the density correlations do
not show anisotropies, while magnetic field fluctuations are anisotropic.
On the basis of our calculations we might expect that the correlation 
may be a bit higher for high $\beta$ than for low $\beta$ plasmas.

{\bf Can viscously damped turbulence explain TSAS?}
The term ``tiny-scale atomic structures''  or TSAS was introduced by
Heiles (1997) to describe the mysterious
H~I absorbing structures on scales from thousands to tens of
AU, discovered by Dieter, Welch \& Romney (1976). Analogs of
TSAS are observed
in NaI and CaII (Meyer \& Blades 1996; Faison \& Goss 2001;
      Andrews, Meyer, \& Lauroesch\ 2001) 
and in molecular gas (Marscher, Moore, \& Bania 1993).

Those structures can be produced by turbulence with a spectrum
substantially more shallow than the Kolmogorov one, e.g. with the
spectrum $E(k)\sim k^{-1}$ (see Deshpande 2000). Simulations
in CLV02b and theoretical calculations
in Lazarian, Vishniac \& Cho (2003) show that the magnetic field
in the viscosity-damped regime of MHD turbulence (see \S\ref{sect_new}) can
produce such a spectrum of magnetic fluctuations. Our calculations
above are indicative that this will translate in the corresponding
shallow spectrum of density. Our calculations   
are applicable on scales from the viscous damping 
scale (determined by equating the energy transfer rate with the 
viscous damping rate; $\sim0.1$ pc in the Warm Neutral Medium with $n$ 
= 0.4 cm$^{-3}$, $T$= 6000 K) to the ion-neutral decoupling scale (the 
scale at which viscous drag on ions becomes comparable to the neutral 
drag; $\ll 0.1$ pc). Below the viscous scale the fluctuations of 
magnetic field obey the damped regime shown in Figure \ref{fig_viscous}b 
and produce 
density fluctuations. For typical Cold Neutral Medium gas, the scale of 
neutral-ion decoupling decreases to $\sim70$ AU, and is less for denser 
gas. TSAS may be created by strongly nonlinear MHD turbulence! 

\subsection{Application of the scaling laws obtained}

Many astrophysical problems require some knowledge of the scaling
properties of turbulence.  Therefore we expect a wide range
of applications of the established scaling relations. Here we discuss how
our understanding of MHD turbulence affects a few selected issues. 

As we mentioned in the introduction, the observations (see CLV03a for
a review) indicate that interstellar spectrum exhibits
Kolmogorov-type scaling $E(k)\sim k^{-5/3}$. On the basis of what we
learned by now (see also Higdon 1983, 
GS95, Lithwick \& Goldreich 2001, CL02) we may conclude that the
MHD spectrum {\it is different } from the spectrum of isotropic Kolmogorov
turbulence. Anisotropy of Alfvenic and slow modes ensures
that the observationally measured spectrum is dominated by perturbations
perpendicular to the magnetic field, i.e. $E(k)\sim k_{\bot}^{-5/3}$.
Admixture of fast modes can alter the spectrum, but observations indicate
that those modes do not dominate the detected signal.   

Whether or not one can use Kolmogorov isotropic scaling for practical
calculations depends on the nature of the problem. For instance, 
Cho \& Lazarian (2002b) use Kolmogorov scaling to calculate 
the spectra of fluctuations arising from synchrotron emission
and of fluctuations of the degree of starlight polarization and obtained
good correspondence with the observed statistics. However, for many problems
turbulence anisotropy is essential.

{\bf Cosmic rays.---}
The propagation of cosmic rays is mainly determined by their interactions
with electromagnetic fluctuations in the interstellar medium. 
For practical calculations it is usually assumed that the turbulence is
{\it isotropic} with a Kolmogorov spectrum (e.g.,  
Schlickeiser \& Miller 1998). 
How should these calculations be modified in view of our findings?
Chandran (2000) and Yan \& Lazarian (2002;
see also Lazarian, Cho \& Yan 2002 for a review)
calculated the  efficiency of cosmic ray scattering by
Alfvenic modes and found that, if the energy injected at
large scales, the scattering by Alfvenic modes is absolutely
negligible. The difference between the calculations in Yan \& Lazarian
(2002), which used the tensor description of Alfven waves obtained
in CLV02a, and the calculations that use Kolmogorov turbulence
was 10 orders of magnitude! 
Yan \& Lazarian (2002; 2003a) identified fast modes as the principal source
of cosmic ray scattering and used the CL02 results to calculate the
scattering rate in low $\beta$ plasma. Damping of those modes, 
however, depends on the value of $\beta$, which entails  
the dependence of cosmic
ray scattering on the temperature and density of the background plasmas.

{\bf Dust dynamics.---}
Turbulence induces relative dust grain motions and leads to 
grain-grain collisions. These collisions
determine grain size distribution, which affects most dust properties,
including absorption and  H$_2$ formation. 
Unfortunately, as in the case of cosmic rays, earlier work appealed
to hydrodynamic turbulence to predict grain relative velocities.
Lazarian \& Yan (2002) and Yan \& Lazarian (2003b)
considered motions of charged grains
in anisotropic MHD turbulence and found that
the direct interaction of the charged grains with turbulent magnetic field
results in a stochastic acceleration that can potentially
provide grains with supersonic velocities.

{\bf Other applications.---}
The obtained scaling laws are essential for understanding the
density fluctuations within HII regions (Lithwick \& Goldreich 2001),
stochastic magnetic reconnection in fully ionized (Lazarian \& Vishniac 1999,
2000) and partially ionized plasma (Lazarian, Vishniac \& Cho 2003),
for energy transfer to electrons in gamma ray bursts and solar flares
(see Lazarian, Petrosian, Yan, \& Cho 2003). This list can be easily extended. 

\section{Summary}

In the paper, we have 
studied statistics of compressible MHD turbulence
for high, intermediate, and low $\beta$ plasmas 
and for different sonic and Alfven Mach numbers. 
For subAlfvenic turbulence we provided the decomposition
of turbulence into Alfven, slow and fast modes.
We have found that the coupling of compressible and incompressible
modes is weak and, contrary to the common belief, the drain of
energy from Alfven to compressible modes is marginal along the
cascade.  
For the cases studied, we have found that GS95
scaling is valid for {\it Alfven modes}:
$$
   \mbox{ Alfv\'{e}n:~}  E^A(k)  \propto k^{-5/3}, 
                        ~~~k_{\|} \propto k_{\perp}^{2/3}. 
$$
{\it Slow modes} also follows the GS95 model for both
high $\beta$ and mildly supersonic low $\beta$ cases:
$$
   \mbox{ Slow:~~~}   E^s(k)  \propto k^{-5/3}, 
                        ~~~k_{\|} \propto k_{\perp}^{2/3}.  
$$
For the highly supersonic low $\beta$ case, the kinetic energy spectrum of 
slow modes tends to be steeper, which may be related
to the formation of shocks.\\
{\it Fast mode} spectra are compatible with
acoustic turbulence scaling relations:
$$
   \mbox{ Fast:~~~}   E^f(k)  \propto k^{-3/2}, 
                        ~\mbox{isotropic spectrum}.   
$$
Our super-Alfvenic turbulence simulations suggest that 
the picture above holds true at sufficiently small scales at which
Alfven speed $V_A$ is larger than the turbulent velocity $v_l$.
This part of our study shows that compressible MHD turbulence is
not a mess. On the contrary, its statistics obeys well determined
universal scaling relations. The importance of these relations
for different branches of Astrophysics is obvious.

Addressing the issue of MHD turbulence damping in partially ionized
gas we showed that the viscosity-damped regime of MHD turbulence below the
viscous damping scale that was reported in CLV02b for incompressible
fluid persists when compressibility is present. The spectrum of
density fluctuations that we obtain may be related to the mysterious
tiny scale structures observed in the ISM.

\section*{Acknowledgments}
We thank Ethan Vishniac, Peter Goldreich, Bill Matthaeus, Chris McKee, 
and Annick Pouquet
for stimulating discussions.
We acknowledge the support of NSF Grant AST-0125544.
This work was partially supported by NCSA
under AST010011N and
utilized the NCSA Origin2000.

\appendix
\section{Mode decomposition}
Let us consider a small perturbation
in the presence of a strong mean magnetic field.
We write density, velocity, pressure, and magnetic field as the sum of 
constant and fluctuating parts:
$\rho \rightarrow \rho_0+\rho$, ${\bf v} \rightarrow {\bf v}_0+{\bf v}$, 
$P \rightarrow P_0+p$,
and ${\bf B} \rightarrow {\bf B}_0+{\bf b}$, respectively.
We assume that ${\bf v}_0=0$ and that
perturbation is small : $\rho << \rho_0$, etc.
Ignoring the second and higher order contributions, we
can rewrite the MHD equations as follows:
\begin{eqnarray}
  \frac{\partial \rho}{\partial t} + \rho_0 \nabla \cdot {\bf v} = 0, \\
  \rho_0 \frac{\partial {\bf v}}{\partial t} + \nabla (a^2 \rho) 
                  - \frac{1}{4\pi} (\nabla \times {\bf b})\times {\bf B}_0=0,
    \\
 \frac{\partial {\bf b}}{\partial t}
      +\nabla \times \left[ {\bf v} \times {\bf B}_0 \right]=0,
\end{eqnarray}
where we assume a polytropic equation of state: 
$p=a^2 \rho$ with $a^2=\gamma p_0/\rho_0$.
We follow arguments in Thompson (1962) to derive magnetosonic waves.
Let ${\bf \xi}({\bf r},t)$ be the displacement vector, so that
$\partial {\bf \xi}/\partial t = {\bf v}$.
Assuming that the displacements vanish at $t=0$, we can integrate
the equations as follows
\begin{eqnarray}
  \rho + \rho_0 \nabla \cdot {\bf \xi} = 0, \\
  \ddot{\bf \xi} = a^2 \nabla (\nabla \cdot {\bf \xi})
   + (\nabla \times {\bf b})\times {\bf B}_0/4\pi\rho_0, 
                                               \label{momen_1stO_intm}\\
  {\bf b} = \nabla \times (\xi \times {\bf B}_0).
\end{eqnarray}
The momentum equation (eq. \ref{momen_1stO_intm}) becomes
\begin{eqnarray}
 \ddot{\bf \xi}=a^2 \nabla (\nabla \cdot {\bf \xi}) +
             \left[ \nabla \times \left( \nabla \times (\xi \times {\bf B}_0) 
                                \right)
             \right] \times {\bf B}_0 /4\pi\rho_0  \nonumber \\
      ~~ =a^2  \nabla (\nabla \cdot {\bf \xi}) +
          \nabla ( B_0^2 \nabla \cdot {\bf \xi} -
         {\bf B}_0 \cdot \nabla {\bf B}_0 \cdot{\bf \xi})/4\pi\rho_0 
                                                               \nonumber \\
     ~~~  - ({\bf B}_0 \cdot \nabla)^2 {\bf \xi}/4\pi\rho_0 
   + \left[{\bf B}_0 ( {\bf B}_0 \cdot \nabla) \nabla \cdot {\bf \xi} 
     \right]/4\pi\rho_0 
\end{eqnarray}
Using $\alpha=a^2/V_A^2=\beta(\gamma/2)$, $V_A=B_0/4\pi\rho_0$, we have
\begin{equation}
\ddot{\bf \xi}/V_A^2-\nabla [ (\alpha+1) \nabla \cdot {\bf \xi}-  
                     (\hat{\bf B}_0 \cdot \nabla)(\hat{\bf B}_0 \cdot {\bf \xi})
                   ]                                    \nonumber \\
  - (\hat{\bf B}_0 \cdot \nabla)^2 {\bf \xi}
  + (\hat{\bf B}_0 \cdot \nabla)(\nabla \cdot {\bf \xi}) \hat{\bf B}_0
  = 0
\end{equation}
In Fourier space
 the equation becomes
\begin{equation}
  \ddot{\bf \xi}/V_A^2 + k \hat{\bf k}[(\alpha +1)k \xi_k - k_{\|} \xi_{\|}]
      + k_{\|}^2 {\bf \xi} - k_{\|} k \xi_{k} \hat{k}_{\|} =0,
  \label{eq_in_FSP}
\end{equation}
where $\xi_k={\bf \xi}\cdot \hat{\bf k}$, 
$\xi_{\|}={\bf \xi}\cdot \hat{\bf k}_{\|}$, $\hat{\bf k}={\bf k}/k$,
and $\hat{\bf k}_{\|}$ is unit vector parallel to ${\bf B}_0$ 
(i.e. $\hat{\bf k}_{\|} = \hat{\bf B}_0$).
Assuming $\ddot{\bf \xi}=-\omega^2 {\bf \xi}=-c^2k^2 {\bf \xi}$,
we can rewrite (\ref{eq_in_FSP}) as
\begin{equation}
   (c^2/V_A^2 -\cos^2 \theta){\bf \xi}
    - [(\alpha+1) \xi_k-\cos \theta \xi_{\|}]\hat{\bf k}
    + \cos \theta \xi_{k} \hat{k}_{\|} =0,
   \label{xi_in_k}
\end{equation}
where $\cos \theta = k_{\|}/k$ and $\theta$ is the angle between
${\bf k}$ and ${\bf B}_0$.

Using $\hat{\bf k}=\sin\theta \hat{\bf k}_{\perp}+\cos\theta \hat{\bf k}_{\|}$,
we get
\begin{eqnarray}
(c^2/V_A^2 -\cos^2 \theta){\bf \xi}
  - [ (\alpha+1)\xi_k-\cos \theta \xi_{\|} ]\sin\theta \hat{\bf k}_{\perp} 
   \nonumber \\ 
  - \{ [ (\alpha+1)\xi_k-\cos \theta \xi_{\|}]\cos\theta 
  - \cos\theta \xi_{k} \} \hat{\bf k}_{\|}=0.
\end{eqnarray}
Writing ${\bf \xi}=\xi_{\perp} \hat{\bf k}_{\perp}
       + \xi_{\|} \hat{\bf k}_{\|}
       + \xi_{\varphi} \hat{\bf \varphi}$,
we get
\begin{eqnarray}
   (c^2/V_A^2 -\cos^2 \theta)\xi_{\perp}-[(\alpha+1)\xi_k-\cos \theta \xi_{\|}]
       \sin\theta =0,      \label{eq_perp}
    \\
    (c^2/V_A^2 -\cos^2 \theta)\xi_{\|}-
    [ \alpha\xi_k-\cos \theta \xi_{\|}]\cos\theta=0,  \label{eq_par}
    \\
     (c^2/V_A^2 -\cos^2 \theta)\xi_{\varphi} = 0.     \label{eq_psi}
\end{eqnarray}

The non-trivial solution of equation (\ref{eq_psi}) is the Alfven wave, whose
dispersion relation is
$\omega/k = V_A \cos\theta$.
The direction of the displacement vector for Alfven wave is parallel to
the azimuthal basis $\hat{\bf \varphi}$:
\begin{equation}
   \hat{\bf \xi}_A = -\hat{\bf \varphi} 
         = \hat{\bf k}_{\perp} \times \hat{\bf k}_{\|}.
\end{equation}

Let us consider solutions of equations (\ref{eq_perp}) and (\ref{eq_par}).
Using $\xi_k=\xi_{\perp}\sin\theta+\xi_{\|}\cos\theta$, we get
\begin{eqnarray}
   (c^2/V_A^2 -\cos^2 \theta)\xi_{\perp}-(\alpha+1)\sin^2\theta \xi_{\perp}-
                              \alpha \cos \theta \sin\theta \xi_{\|}=0,
    \\
    (c^2/V_A^2 -\cos^2 \theta)\xi_{\|}-
   \alpha\sin\theta \cos\theta \xi_{\perp}-
    (\alpha-1)\cos^2 \theta \xi_{\|}=0.
\end{eqnarray}
Rearranging these, we get
\begin{eqnarray}
   (c^2/V_A^2 -\alpha \sin^2\theta-1)\xi_{\perp}-
                              \alpha \cos\theta \sin\theta \xi_{\|}=0,
                   \label{kpar_and_kperp1}
    \\
    (c^2/V_A^2 -\alpha \cos^2 \theta)\xi_{\|}-
   \alpha \sin\theta \cos\theta \xi_{\perp}=0.    \label{kpar_and_kperp2}
\end{eqnarray}
Combining these two, we get
\begin{eqnarray}
  (c^2/V_A^2 -\alpha \sin^2\theta-1)
  (c^2/V_A^2 -\alpha \cos^2 \theta)
  \nonumber \\
   = \alpha^2 \sin^2\theta \cos^2\theta .
\end{eqnarray}
Therefore, the dispersion relation is
\begin{equation}
   c^4/V_A^4 - (1+\alpha) c^2/V_A^2 + \alpha \cos^2\theta = 0.
\end{equation}
The roots of the equation are
\begin{equation}
   c_{f,s}^2= \frac{1}{2} V_A^2 \left[ (1+\alpha) \pm 
               \sqrt{ (1+\alpha)^2 - 4\alpha \cos^2\theta } \right], 
        \label{c_sf}
\end{equation}
where subscripts `f' and 's' stand for `fast' and 'slow' waves, respectively.

We can write
\begin{eqnarray}
   {\bf \xi}=\xi_{\|} \hat{\bf k}_{\|} + \xi_{\perp} \hat{\bf k}_{\perp} 
   \propto \left[\frac{\xi_{\|}k_{\perp}}{\xi_{\perp}k_{\|}}\right]
            k_{\|} \hat{\bf k}_{\|} 
     + 
      k_{\perp} \hat{\bf k}_{\perp}.
\end{eqnarray}
Plugging eq. (\ref{c_sf}) into eq.~(\ref{kpar_and_kperp1}) and 
(\ref{kpar_and_kperp2}), we get
\begin{eqnarray}
   \left[ \frac{1+\alpha}{2} \pm \frac{\sqrt{D}}{2} 
                -\alpha \sin^2\theta -1               \right] \xi_{\perp}
     = \alpha \cos\theta \sin\theta \xi_{\|}, \\
   \left[ \frac{1+\alpha}{2}\pm \frac{\sqrt{D}}{2} 
                -\alpha \cos^2\theta \right] \xi_{\|}
     = \alpha \cos\theta \sin\theta \xi_{\perp},
\end{eqnarray}
where $D=(1+\alpha)^2-4{\alpha} \cos^2{\theta}$.
Using $k_{\|}=k\cos\theta$ and 
$k_{\perp}=k\cos\theta$, we get
\begin{eqnarray}
   \left[ \frac{-1+\alpha}{2}\pm \frac{\sqrt{D}}{2} 
   \right] \xi_{\perp}k_{\|}
                -\alpha \sin^2\theta   \xi_{\perp}k_{\|}
     = \alpha \cos^2\theta \xi_{\|}k_{\perp}, \\
   \left[ \frac{1+\alpha}{2}\pm \frac{\sqrt{D}}{2} 
   \right]\xi_{\|}k_{\perp}
                -\alpha \cos^2\theta \xi_{\|}k_{\perp}
     = \alpha  \sin^2\theta \xi_{\perp}k_{\|}.
\end{eqnarray}
Arranging these, we get
\begin{equation}
    \frac{ \xi_{\|}k_{\perp} }{ \xi_{\perp}k_{\|} }
    = \frac{ -1 + \alpha \pm \sqrt{D} }{ 1+\alpha \pm \sqrt{D} },
\end{equation}
where the upper signs are for fast mode and the lower signs for slow mode.
Therefore, we get
\begin{eqnarray}
   \hat{\bf \xi}_s \propto 
     ( -1 + \alpha - \sqrt{D} )
            k_{\|} \hat{\bf k}_{\|} 
     + 
     ( 1+\alpha - \sqrt{D} ) k_{\perp} \hat{\bf k}_{\perp},
  \label{eq_xis_new}
\\
   \hat{\bf \xi}_f \propto 
     ( -1 + \alpha + \sqrt{D} )
            k_{\|} \hat{\bf k}_{\|} 
     + 
     ( 1+\alpha + \sqrt{D} ) k_{\perp} \hat{\bf k}_{\perp}. 
   \label{eq_xif_new}
\end{eqnarray}
The slow basis $\hat{\bf \xi}_s$ lies between $\hat{\bf k}_{\|}$ and
$-\hat{\bf \theta}$.
The slow basis $\hat{\bf \xi}_f$ lies between $\hat{\bf k}_{\perp}$ and
$\hat{\bf k}$ (Fig.~\ref{fig_separation}).
Here overall sign of $\hat{\bf \xi}_s$ and $\hat{\bf \xi}_f$ is not important.

When $\alpha \rightarrow 0$,
equations (\ref{eq_xif_new}) and (\ref{eq_xis_new}) becomes
\begin{eqnarray}
   \hat{\bf \xi}_s \approx  \hat{\bf k}_{\|}
     -(\alpha \sin\theta \cos\theta)\hat{\bf k}_{\perp}, \label{xis_lowbeta}
\\
   \hat{\bf \xi}_f \approx  (\alpha \sin\theta \cos\theta) \hat{\bf k}_{\|}
                         +\hat{\bf k}_{\perp}.    \label{xif_lowbeta}     
\end{eqnarray}
In this limit, $\hat{\bf \xi}_s$ is mostly proportional to $\hat{\bf k}_{\|}$
and $\hat{\bf \xi}_f$ to $\hat{\bf k}_{\perp}$.
When $\alpha \rightarrow \infty$,
equations (\ref{eq_xif_new}) and (\ref{eq_xis_new}) becomes
\begin{eqnarray}
   \hat{\bf \xi}_s \approx  -\hat{\bf \theta}
             +(\sin\theta \cos\theta/\alpha)\hat{\bf k}, \label{xis_highbeta}
\\
   \hat{\bf \xi}_f \approx
                 (\sin\theta \cos\theta/\alpha) \hat{\bf \theta}
                                   +\hat{\bf k}.    \label{xif_highbeta}
\end{eqnarray}
When $\alpha = \infty$, slow modes are called {\it pseudo}-Alfvenic modes.

We can obtain slow and fast velocity component 
by projecting Fourier velocity component 
${\bf v}_{\bf k}$ onto $\hat{\bf \xi}_s$ and $\hat{\bf \xi}_f$, respectively.

{}To separate slow and fast magnetic modes, 
we assume the linearized continuity equation
($\omega \rho_k = \rho_0 {\bf k} \cdot  {\bf v}_k$) and
the induction equation
($\omega {\bf b}_k = {\bf k} \times ({\bf B}_0 \times {\bf v}_k)$)
are {\it statistically} true.
{}From these, we get Fourier components of density
and {\it non-Alfv\'{e}nic} magnetic field:
\begin{eqnarray}
\rho_k 
       &=&(\rho_0 \Delta v_{k,s}/c_s) \hat{\bf k}\cdot \hat{\bf \xi}_s
      +(\rho_0 \Delta v_{k,f}/c_f) \hat{\bf k}\cdot \hat{\bf \xi}_f
        \nonumber \\
       &\equiv&\rho_{k,s}+\rho_{k,f},   \label{eq_rho}   \\
b_k    &=&
       (B_0 \Delta v_{k,s}/c_s) |\hat{\bf B}_0\times \hat{\bf \xi}_s|
      +(B_0 \Delta v_{k,f}/c_f) |\hat{\bf B}_0\times \hat{\bf \xi}_f|
     \nonumber \\
    &\equiv&b_{k,s}+b_{k,f},   \label{eq_b1}  \\
    &=& \rho_{k,s} (B_0/\rho_0)
      (|\hat{\bf B}_0\times \hat{\bf \xi}_s|/\hat{\bf k}\cdot \hat{\bf \xi}_s)
                                     \nonumber   \\
      &+& \rho_{k,f} (B_0/\rho_0)
      (|\hat{\bf B}_0\times \hat{\bf \xi}_f|/\hat{\bf k}\cdot \hat{\bf \xi}_f),
             \label{eq_b2}
\end{eqnarray}
where $\Delta v_k \propto v_k^+-v_k^-$ (superscripts `+' and `-'
represent opposite directions of wave propagation) 
and subscripts `s' and `f' stand for `slow' and
`fast' modes, respectively. 
{}From equations (\ref{eq_rho}), (\ref{eq_b1}), and (\ref{eq_b2}),
we can obtain $\rho_{k,s}$, $\rho_{k,f}$, $b_{k,s}$, and $b_{k,f}$
 in Fourier space.

\section{Scale-dependent anisotropy}
\begin{figure*}
  \includegraphics[width=0.48\textwidth]{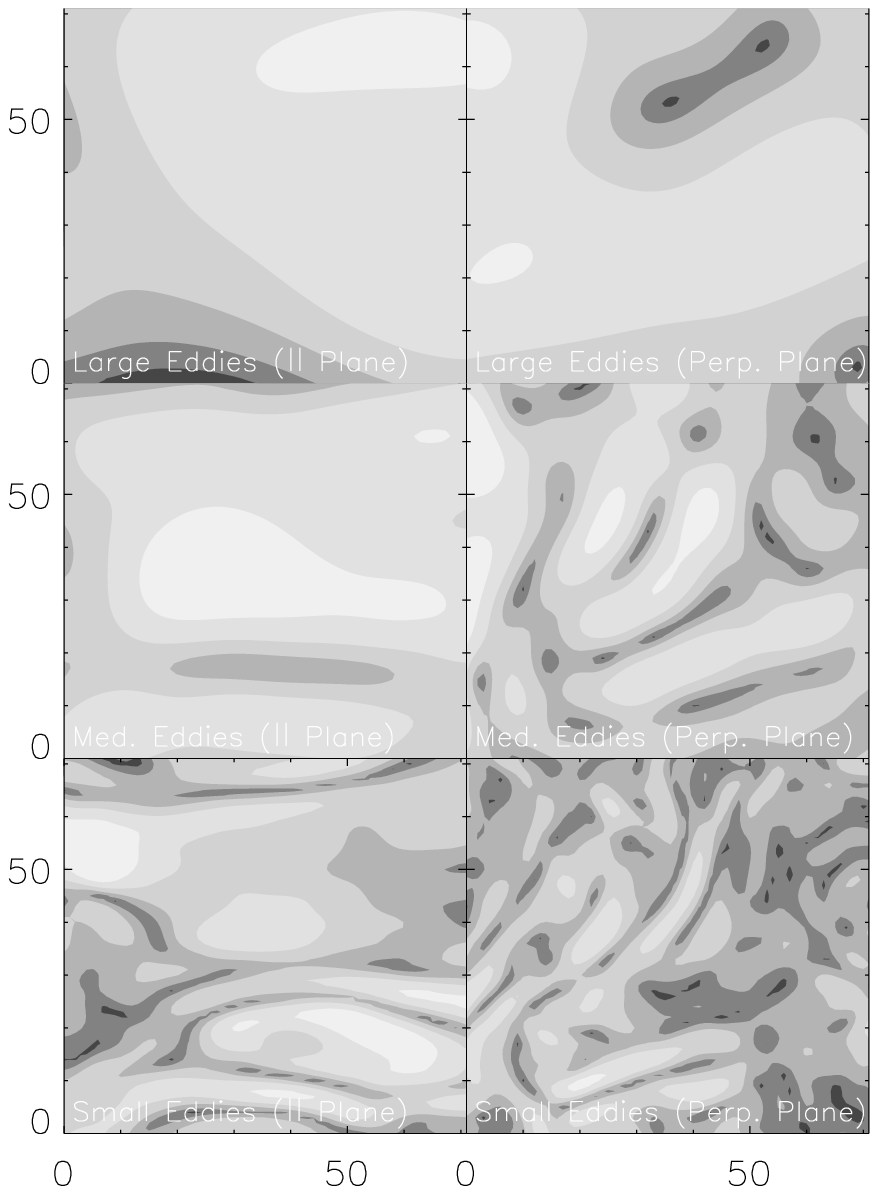}
\hfill
  \includegraphics[width=0.48\textwidth]{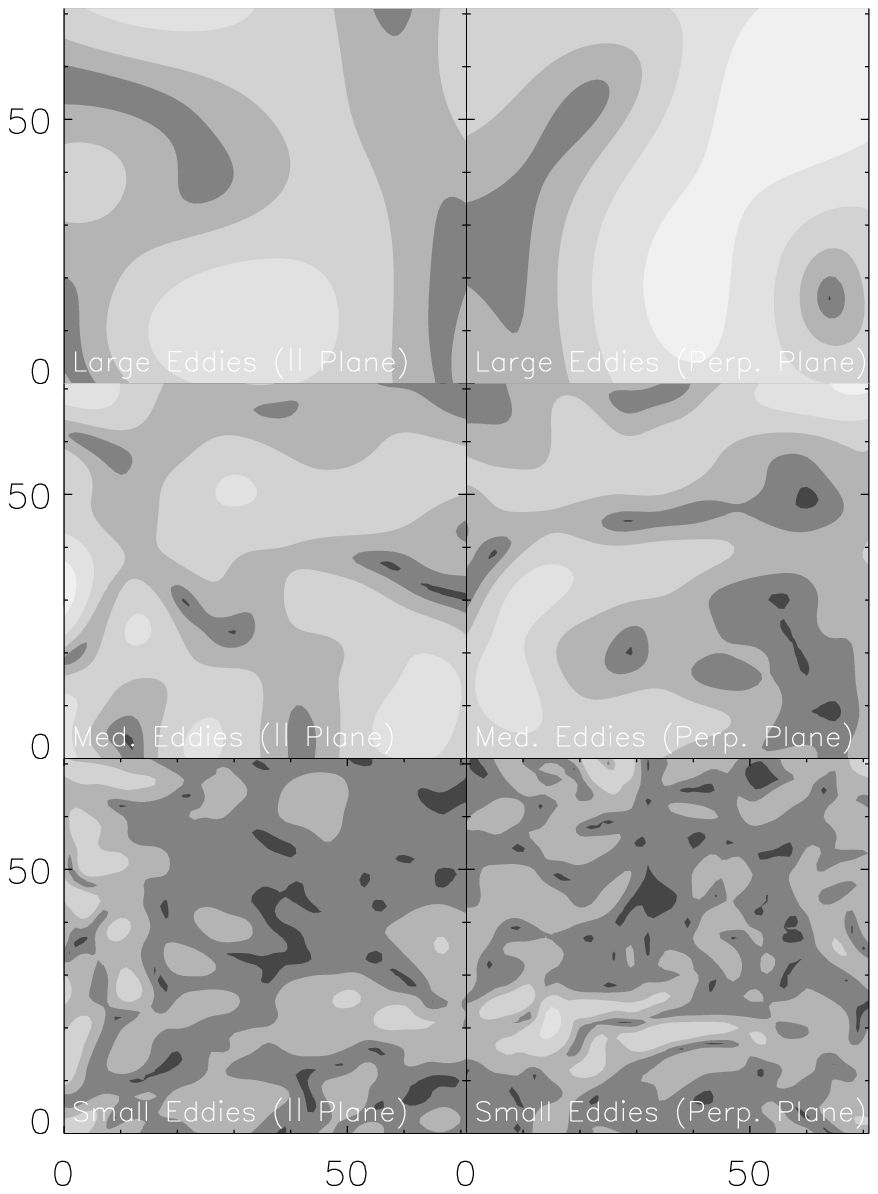}
  \caption{
      Anisotropy as a function of scale.
    ({\it Left}) Alfven mode velocity show scale-dependent anisotropy.
    ({\it Right}) Fast mode velocity show isotropy. Only part of the data cube
     is shown. Lighter tones are for larger $|{\bf v}|$.
    Magnetic field show similar behaviors.
}
\label{fig_6figs}
\end{figure*}

{}Fig.~\ref{fig_6figs}a shows the shapes of Alfven eddies of different sizes.
Left 3 panels show an increased anisotropy 
as we move from the top (large eddies) to the bottom (small eddies).
The horizontal axes of the left panels are parallel to ${\bf B_0}$.
Structures in the perpendicular plane (right panels)
do not show a systematic 
elongation. 
However, Fig.~\ref{fig_6figs}b shows that
velocity of fast modes exhibit isotropy.
Data are from a simulation with $216^3$ grid points, $M_s=2.3$, and $\beta=0.2$.

\end{document}